\documentclass[a4paper,11pt]{article}

\usepackage[T1]{fontenc}
\usepackage{textcomp}

\usepackage[ngerman,english]{babel}

\usepackage[usenames,dvipsnames]{color}

\usepackage{amssymb}
\usepackage{amsthm}
\usepackage{amsmath}

\usepackage{calc}

\usepackage[mathscr]{eucal}
\usepackage{graphicx}
\usepackage{psfrag}
\usepackage{subfigure}

\graphicspath{{.}{.}}

\usepackage{fancyhdr}

\usepackage{bbm}     
\usepackage{dsfont}  

\usepackage{bm}

\usepackage{pifont}

\usepackage{fancyhdr}

\usepackage{tensor}

\usepackage{stmaryrd} 

\usepackage{undertilde}

\usepackage{varioref}

\usepackage{marvosym}

\usepackage{upgreek}

\usepackage{cancel}

\usepackage{chemarrow}

\renewcommand{\sectionmark}[1]%
        {\markboth%
                {}%
                {{\rm\thesection}\quad{\sc #1}}}

\fancypagestyle{plain}{%
\fancyhf{}
\fancyfoot[LE,RO]{\bfseries \thepage}

}

\newcommand{\proofend}{\hspace*{\fill}\rule{0.2cm}{0.2cm}}

\newcommand{\textfrac}[2]{{\textstyle{\frac{#1}{#2}}}}

\newcommand{\threemetric}{g}
\newcommand{\fourmetric}{{}^4\mathbf{g}}
\newcommand{\metricS}{\mathbf{g}_{[S^2]}}
\newcommand{\twoDelta}{\tilde{D}}

\newcommand{\rem}{r_{\mathrm{m}}}
\newcommand{\rk}{r_{\mathrm{k}}}
\newcommand{\Kk}{K_{\mathrm{k}}}
\newcommand{\Kem}{K_{\mathrm{m}}}
\newcommand{\Cem}{C_{\mathrm{m}}}
\newcommand{\Tem}{\mathcal{T}_{\mathrm{m}}}
\newcommand{\tauem}{\tau_{\mathrm{m}}}

\DeclareMathOperator{\tr}{tr}

\DeclareMathOperator{\arsinh}{arsinh}

\theoremstyle{plain}
\newtheorem{theorem}{Theorem}[section]

\newtheorem{proposition}[theorem]{Proposition}

\newtheorem*{definition}{Definition}

\theoremstyle{remark}

\newtheorem*{remark}{Remark}

\begin{document}

\title{\huge \sc Constant mean curvature slicings of Kantowski-Sachs spacetimes}

\author{\\
{\Large\sc J.\ Mark Heinzle}\thanks{Electronic address:  {\tt Mark.Heinzle@univie.ac.at}} \\[0.5ex] 
University of Vienna, Faculty of Physics, Gravitational Physics \\
Boltzmanngasse 5, 1090 Vienna, Austria \\[2ex] }

\date{}
\maketitle
\begin{abstract}
We investigate existence, uniqueness, and the asymptotic properties
of constant mean curvature (CMC) slicings in vacuum
Kantowski-Sachs spacetimes with positive cosmological constant.
Since these spacetimes violate the strong energy condition, most of the
general theorems on CMC slicings do not apply.
Although there are in fact Kantowski-Sachs spacetimes with a unique CMC foliation 
or CMC time function, we prove that there also exist
Kantowski-Sachs spacetimes with an arbitrary number
of (families of) CMC slicings. The properties of these slicings
are analyzed in some detail.

\end{abstract}

\vspace{2cm}

\begin{center}
Keywords: \begin{minipage}[t]{10cm}
Constant mean curvature -- CMC slicing -- CMC foliation -- cosmological constant -- Kantowski-Sachs -- Schwarzschild-de Sitter
\end{minipage}
\end{center}

\vfill
\newpage

\section{Introduction}
\label{introduction}

In the context of spacetimes that admit spatially compact spacelike Cauchy hypersurfaces,
hypersurfaces of constant mean curvature (CMC) and slicings by such hypersurfaces
have attracted a lot of attention.
One major motivation for studying such slicings rests on the fact that
CMC foliations give rise to a geometrically distinguished time-function
on the spacetime. CMC slicings thus appear prominently whenever cosmological spacetimes are
discussed, when the strong cosmic censorship conjecture is investigated,
and they also play an important role in numerical studies.

There exist several well-established properties of CMC hypersurfaces in
cosmological spacetimes satisfying the timelike convergence condition (which is identical to the strong energy condition
and states that the contraction of the Ricci tensor with any timelike vector is non-negative).
In particular, uniqueness of CMC hypersurfaces is well-understood:
In a cosmological spacetime satisfying the timelike convergence condition, 
for any given constant $K_0 \neq 0$,
there exists at most one compact CMC hypersurface with mean curvature $\tr k = K_0$;
the proof in~\cite{Marsden/Tipler:1980} is based on the techniques of~\cite{Brill/Flaherty:1976}.
Modulo some subtleties, an analogous result holds for maximal hypersurfaces (where $\tr k=0$).
A consequence of the uniqueness result is that any Killing vector the spacetime possesses
must be tangential to the CMC hypersurface.
The existence of compact CMC hypersurfaces (and foliations thereof) is a more delicate issue.
(If there exists one compact CMC hypersurface, then there exists a  
foliation of a neighborhood of that hypersurface by compact CMC hypersurfaces.
This follows from standard arguments if $\tr k \neq 0$; 
if the initial hypersurface is maximal, see~\cite{Bartnik:1988}.)
If the spacetime possesses barriers, i.e., compact hypersurfaces $\mathcal{S}_1$
and $\mathcal{S}_2$, $\mathcal{S}_2$ in the chronological future of $\mathcal{S}_1$, 
with mean curvature $\tr k_1$ and $\tr k_2$ such that 
$\sup_{\mathcal{S}_1} \tr k_1 < \inf_{\mathcal{S}_2} \tr k_2$, then there exists
a foliation of compact CMC hypersurfaces between $\mathcal{S}_1$ and $\mathcal{S}_2$
where the mean curvature assumes each value between $\sup_{\mathcal{S}_1} \tr k_1$ and $\inf_{\mathcal{S}_2} \tr k_2$, see~\cite{Gerhardt:1983}.
A spacetime with crushing singularities is globally foliated by compact CMC hypersurfaces~\cite{Gerhardt:1983,Ecker/Huisken:1991}.

It has been conjectured that every maximal globally hyperbolic 
vacuum cosmological spacetime (satisfying the
timelike convergence condition) with 
compact CMC hypersurface admits a global CMC foliation,
where the mean curvature ranges over $(-\infty,\infty)$ or $(-\infty,0)$
depending on the Yamabe class of the CMC slice~\cite{Andersson:2004}.
In certain special cases the conjecture has been shown to be true, e.g.,
for locally spatially homogeneous spacetimes, and for spherically symmetric
spacetimes with spatial topology $S^1\times S^2$, see \cite[Sec.~2]{Rendall:1996a} for a review 
and~\cite{Burnett/Rendall:1996} and references therein.
However, the conjecture is not true in general. 
There exist examples of spacetimes with a CMC foliation that is
not global~\cite{Isenberg/Rendall:1998} and spacetimes that 
do not possess compact CMC hypersurfaces at all~\cite{Bartnik:1988}.
The failure of the conjecture cannot be attributed to the presence of matter,
since there exist vacuum cosmological spacetimes without compact 
CMC hypersurfaces as well~\cite{Chrusciel/Isenberg/Pollack:2004}.

Without the assumption of the timelike convergence condition, 
fewer results are available:
Uniqueness of a globally defined CMC time function with compact level sets
still holds~\cite{Barbot/Beguin/Zeghib:2007} (and the associated
CMC foliation is the unique CMC foliation).
In~\cite{Gerhardt:2006} another uniqueness result is proved
under the assumption that there exists $\Lambda>0$ such that $R_{\mu\nu} v^\mu v^\nu \geq -\Lambda$ $\forall v^\mu v_\mu =-1$:
For any given constant $|K_0| > \sqrt{3\Lambda}$,
there exists at most one compact CMC hypersurface with mean curvature $\tr k = K_0$.
Existence of a CMC foliation can be deduced from the existence of barriers~\cite{Gerhardt:2006}:
If the spacetime possesses a future crushing singularity (future mean curvature barrier),
then a future end of the spacetime admits a CMC foliation, where the range of the mean
curvature is a subset of the interval $(\sqrt{3 \Lambda},\infty)$.
(The proofs are based on the crucial fact that 
$k_{i j} k^{i j} + R_{\mu\nu} v^\mu v^\nu \geq (\tr k)^2/3 - \Lambda >0$ for
hypersurfaces with $\tr k > \sqrt{3\Lambda}$. 
The expression $k_{i j} k^{i j} + R_{\mu\nu} v^\mu v^\nu$ appears in the lapse equation, 
see~\eqref{lapse}.)

Since it is as yet unclear which general properties are to be expected
of CMC foliations in spacetimes violating the timelike convergence condition,
the detailed study of CMC hypersurfaces in explicit classes of spacetimes
comes into focus.
In~\cite{Beig/Heinzle:2005}, a class of spatially compact Kottler-Schwarschild-de Sitter spacetimes 
has been investigated in detail;
these are vacuum solutions of the Einstein equations with positive cosmological constant $\Lambda$,
which causes the timelike convergence condition to be violated, 
see Section~\ref{KSKSdS} for the definition.
It is found that the spacetimes of this class do not contain compact CMC hypersurfaces 
with $|\tr k|\geq \sqrt{3\Lambda}$. Each spacetime contains a
unique family of slicings of compact CMC hypersurfaces,
where the mean curvature ranges between $(-\sqrt{3\Lambda},\sqrt{3\Lambda})$.
However, the slicings do not cover the entire spacetime, and,
in general, the slicings are foliations only during some time of their evolution.

In this paper we investigate the properties of 
CMC slicings in vacuum \textit{Kantowski-Sachs} models
with cosmological constant $\Lambda > 0$, which we introduce in some detail in section~\ref{KSKSdS}.
Let us summarize the main results we prove in this work:
There are two (of eight) classes of Kantowski-Sachs spacetimes for which 
there exists a global CMC time function and thus a unique global CMC foliation.
For the remaining classes of Kantowski-Sachs models this is not true in general.
We show that, on the one hand, there are Kantowski-Sachs spacetimes with
a unique global CMC foliation (which does, however, not define a time function);
on the other hand, there are Kantowski-Sachs spacetimes containing
(arbitrarily) many families of CMC slicings,
where the range of $\tr k$ is $(\Kem,\sqrt{3\Lambda})$ (with \mbox{$\Kem > \sqrt{\Lambda}$}).
In the future `end' of the spacetime, these slicings are foliations.

Note that, in this paper, a \textit{slicing} denotes a smooth family of 
smooth (spacelike) hypersurfaces. A
parametrization of a slicing is a smooth map $\Psi: I\times \Sigma \rightarrow \text{spacetime}$,
where $\Sigma$ is a \mbox{3-manifold} and $I \subseteq \mathbb{R}$, such that for all $\tau\in I$,
$\Psi(\tau,\cdot)$ is an embedding. We require that $\Psi(\tau, \Sigma)$ is a hypersurface of the
slicing for all $\tau\in I$, i.e., by the parametrization of the slicing the hypersurfaces are
represented as level sets $\tau =\mathrm{const}$.
Note that a slicing is a \textit{foliation} iff
the map $\Psi$ is a diffeomorphism onto its image.

Section~\ref{KSKSdS} of the paper is perhaps a (short) paper in its own right.
It contains a concise discussion of the (vacuum, $\Lambda >0$) Kantowski-Sachs models 
with a focus on the geometric interpretation of these models
in terms of the Kottler-Schwarzschild-de Sitter metric.
In section~\ref{data} we analyze spherically symmetric CMC initial data sets,
which are embedded into Kantowski-Sachs (and Kottler-Schwarzschild-de Sitter) spacetimes in
the subsequent section.
Finally, sections~\ref{Sec:CdST} and~\ref{properties} contain the analysis and the results
on CMC slicings and foliations in Kantowski-Sachs spacetimes.

\section{Kantowski-Sachs and Kottler-Schwarzschild-de Sitter}
\label{KSKSdS}

\subsection{Kantowski-Sachs spacetimes}

The Kantowski-Sachs models
are a class of locally rotationally symmetric (LRS) and spatially 
homogeneous (SH) spacetimes~\cite{Collins:1977,Kantowski:1998}.
The defining property of these models is the existence of 
a four-dimensional isometry group,
whose orbits are three-dimensional spacelike hypersurfaces, where, however, 
there does not exist any three-dimensional subgroup
that acts simply transitively on these orbits. 
The metric of a Kantowski-Sachs spacetime is
\begin{equation}\label{KSmetric}
\fourmetric = - d \hat{t}^{\,2} + \hat{g}_{11}(\hat{t})\, d \hat{r}^2 + \hat{g}_{22}(\hat{t})\: \metricS\:,
\end{equation}
where $\metricS$ is the standard metric on the $2$-sphere.
Assuming that the coordinate $\hat{r}$ ranges in $S^1$, the spatial topology is
$S^1 \times S^2$ and thus compact.

In this paper we consider Kantowski-Sachs spacetimes $(M, \fourmetric)$ that satisfy 
the Einstein equations with a positive cosmological constant $\Lambda$.
The qualitative dynamics of these models has been partly investigated in~\cite{Weber:1984, Moniz:1993};
the paper~\cite{Goliath/Ellis:1999} contains a fairly complete picture.
This subsection is a concise compendium.

Every (vacuum) Kantowski-Sachs model (with $\Lambda >0$) is 
described by a metric 
\begin{equation}\label{KSmetricp}
\fourmetric = - D^{-2}(\tau) \, d \tau^{2} + g_{1 1}(\tau)\, d \hat{r}^2 + g_{2 2}(\tau)\: \metricS\:,
\tag{\ref{KSmetric}${}^\prime$}
\end{equation}
whose components $D$, $g_{1 1}$, and $g_{2 2}$ are determined 
by the equations
\begin{equation}\label{g11}
D = \sqrt{\frac{\Lambda}{3}} \: \frac{1}{\sqrt{1- \Sigma_+^2}} \:,\quad
\partial_\tau g_{1 1} = -2 g_{1 1}\, (-H_D + 2 \Sigma_+ ) \:,\quad
g_{2 2} = \frac{1}{\Lambda}\: \frac{1 - \Sigma_+^2}{1 - H_D^2} \:,
\end{equation}
where $(-\infty,\infty) \ni \tau \mapsto (H_D,\Sigma_+)(\tau)$
is a `seed solution', i.e., a solution of the system of equations
\begin{subequations}\label{reddynsys}
\begin{align}
\partial_\tau H_D & = -(1-H_D^2) (-1 + 3 \Sigma_+^2 - H_D \Sigma_+ )  \:,\\[0.5ex]
\partial_\tau \Sigma_+ & = - (1-\Sigma_+^2) (1 + 3 H_D \Sigma_+ - H_D^2) \:,
\end{align}
\end{subequations}
which is a dynamical system on the relatively compact state space $(0,1) \times (0,1)$.

We refer to~\cite{Goliath/Ellis:1999} or~\cite{Calogero/Heinzle:2010} 
for a derivation of the `reduced dynamical system'~\eqref{reddynsys}.
(We use the intuitive notation of~\cite{Calogero/Heinzle:2010};
to compare with~\cite{Goliath/Ellis:1999} one sets $H_D = Q_0$, $\Sigma_+ = Q_+$.)
Let us merely note 
that $H_D$ and $\Sigma_+$ are normalized versions of
the Hubble expansion $H$ 
and the shear variable $\sigma_+$,
\begin{subequations}\label{HDSig+}
\begin{alignat}{3}
H_D & = H/D & & = H\: \big(H^2 + \textfrac{1}{3}\,g_{2 2}^{-1}\big)^{-1/2}  & & = 
\textfrac{1}{3}\,\tr k \:\, \Big(\textfrac{1}{9}\, (\tr k)^2 + \textfrac{1}{3}\,g_{2 2}^{-1}\Big)^{-1/2}\:, \\[0.5ex]
\Sigma_+ & = \sigma_+/D & & = \sigma_+\, \big(H^2 + \textfrac{1}{3}\,g_{2 2}^{-1}\big)^{-1/2} & & = 
-\textfrac{1}{3} \,( \tensor{k}{^1_1} - \tensor{k}{^2_2})\, \Big(\textfrac{1}{9}\, (\tr k)^2 + \textfrac{1}{3}\,g_{2 2}^{-1}\Big)^{-1/2}\:.
\end{alignat}
\end{subequations}
In~\eqref{HDSig+}, $k_{i j}$ is the extrinsic curvature of the hypersurfaces of homogeneity 
(where we use the convention $\partial_{\hat{t}}\, g_{i j} = 2 k_{i j}$) 
and $\tr k =  \tensor{k}{^1_1} + 2 \tensor{k}{^2_2}$.

\begin{figure}[htp]
  \centering
  \includegraphics[width=1\textwidth]{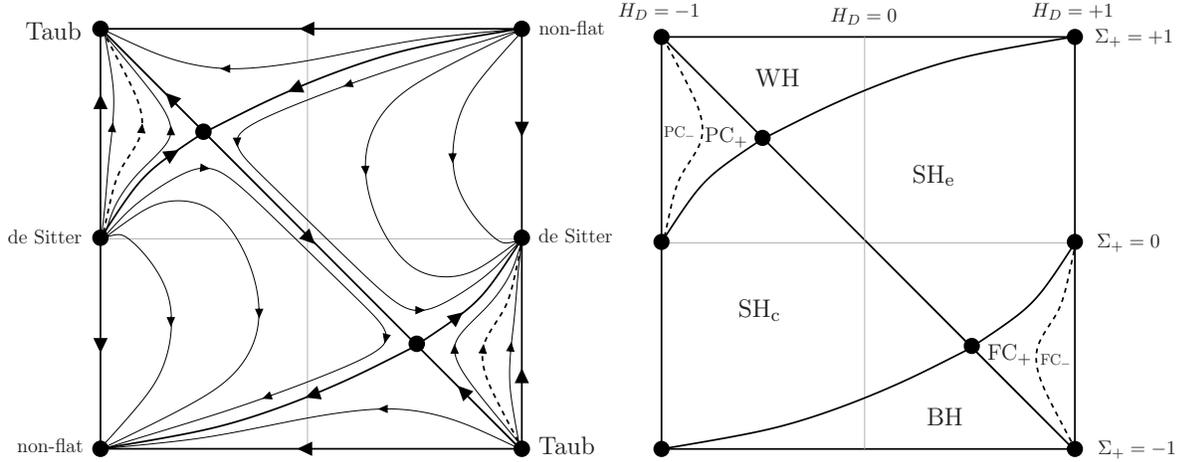}
  \caption{{\small The figure on the l.h.\ side gives the 
      flow of the system~\eqref{reddynsys} on $[-1,1]\times [-1,1] \ni (H_D, \Sigma_+)$.
      The labeling of the fixed points refers to the asymptotic behavior of solutions
      that converge to these points.
      The figure on the r.h.\ side shows a partition of the state space into
      eight invariant domains of positive measure: BH (black hole), WH (white hole),
      FC${}_+$ (future cosmological region, positive mass), 
      FC${}_-$ (future cosmological region, negative mass), 
      PC${}_+$ (past cosmological region, positive mass), 
      PC${}_-$ (past cosmological region, negative mass), 
      SH${}_{\mathrm{e}}$ (spatially homogeneous, mainly expanding),
      SH${}_{\mathrm{c}}$ (spatially homogeneous, mainly contracting). 
      This nomenclature will become clear in the context of the discussion of the Kottler-Schwarzschild-de Sitter spacetimes.}}
  \label{KSflow}
\end{figure}

The dynamical system~\eqref{reddynsys} encodes the entire dynamics of (vacuum, $\Lambda >0$) Kantowski-Sachs models.
We omit the analysis of~\eqref{reddynsys} and
content ourselves with the crucial facts:
First, the system is smoothly extendible to $[-1,1] \times [-1,1]$;
second, the straight line \mbox{$H_D + \Sigma_+ = 0$} is an invariant set;
third, the ellipse $H_D^2 + 3 \Sigma_+^2 = 1$, in combination with the line $H_D + \Sigma_+ = 0$,
defines future and past invariant subsets of the state space. 
The result of the analysis is a complete and rigorous picture of the flow of the dynamical system~\eqref{reddynsys}.
This picture is given in Fig.~\ref{KSflow}.

A solution that converges to a fixed point labeled `Taub' in Fig.~\ref{KSflow}
exhibits `Taub asymptotics', which means 
$\hat{g}_{1 1}(\hat{t}) \sim (\hat{t}-\hat{t}_\pm)^2$ and $\hat{g}_{2 2}(\hat{t}) \sim \mathrm{const}$
as $\hat{t}\searrow \hat{t}_-$ or $\hat{t} \nearrow \hat{t}_+$, respectively,
where $\hat{t}_\pm$ represents an initial/final singularity (in cosmological time $\hat{t}$). 
The label `non-flat' refers to asymptotic behavior governed by the
non-flat LRS Kasner solution; for a solution converging to one of these points
we have $\hat{g}_{1 1}(\hat{t}) \sim (\hat{t}-\hat{t}_\pm)^{-2/3}$ and $\hat{g}_{2 2}(\hat{t}) \sim (\hat{t}-\hat{t}_\pm)^{4/3}$
as $\hat{t}\searrow \hat{t}_-$ or $\hat{t} \nearrow \hat{t}_+$, respectively.
Finally, de Sitter asymptotics corresponds to
$\hat{g}_{1 1}(\hat{t}) \sim \exp(\pm\sqrt{\Lambda/3}\:\hat{t})$
and $\hat{g}_{2 2}(\hat{t}) \sim \exp(\pm\sqrt{\Lambda/3}\:\hat{t})$ as $\hat{t}\rightarrow \pm \infty$.

The orbits converging to the two interior fixed points form a net of separatrices, 
which yields a partition of the state space into six invariant subsets.
In addition, there are two special orbits of~\eqref{reddynsys}, which
are depicted by dashed lines in Fig.~\ref{KSflow}. These orbits correspond
to the models described in~\cite{Gron:1986}, which are
the unique Kantowski-Sachs models that are isotropic~\cite{Torrence/Couch:1988}. We thereby obtain
a partition of the state space into eight domains of positive measure,
whose significance will be discussed in the context of  the Kottler-Schwarzschild-de Sitter spacetimes.

We conclude this section with a brief discussion of the expansion of Kantowski-Sachs models.
From~\eqref{g11} and~\eqref{HDSig+} we have
\begin{equation}\label{expans}
H = \sqrt{\frac{\Lambda}{3}} \: \frac{H_D}{\sqrt{1-\Sigma_+^2}} \quad\text{and}\quad
\tr k = 3 H = \sqrt{3 \Lambda} \: \frac{H_D}{\sqrt{1-\Sigma_+^2}}\:,
\end{equation}
which makes it straightforward to obtain the expansion of models from Fig.~\ref{KSflow}.
It is interesting to note that there exist models (in SH${}_{\mathrm{e}}$) that exhibit
an intermediate phase of contraction.
Furthermore, analysis of the equation $\partial_\tau H = D ( 1 - H_D^2 - 3\Sigma_+^2)$
shows that the time-derivative of $H$ can change sign at most once.
All of the five classes of initially expanding models (BH, WH, FC${}_\pm$, SH${}_{\mathrm{e}}$) 
are initially decelerating; however, for three classes (FC${}_\pm$, SH${}_{\mathrm{e}}$),
initial deceleration is followed by late-time acceleration (associated with de Sitter asymptotics).

\subsection{Kottler-Schwarzschild-de Sitter spacetimes}
\label{KSdSsubsec}

The Kottler-Schwarzschild-de Sitter metric~\cite{Kottler:1918} reads
\begin{equation}\label{KSSdSmetric}
\fourmetric = -V d t^2 + V^{-1} d r^2 + r^2 \: \metricS \,\quad\text{with}\quad V= V(r) = 1-\frac{2 M}{r} -\frac{\Lambda r^2}{3}\:.
\end{equation}
We assume a cosmological constant $\Lambda>0$.
If the constant $M > 0$ satisfies $9 M^2 \Lambda <1$, 
then there exists a non-trivial interval $(r_{\mathrm{b}}, r_{\mathrm{c}})$
such that $V(r_{\mathrm{b}}) = V(r_{\mathrm{c}}) =0$ and $V(r) > 0$ in $(r_{\mathrm{b}}, r_{\mathrm{c}})$.
The region $r_{\mathrm{b}} < r < r_{\mathrm{c}}$ is a static region of the spacetime~\eqref{KSSdSmetric}
with Killing vector $\xi = \partial_t$.
It is straightforward to see that
$2 M < r_{\mathrm{b}} < 3 M < 1/\sqrt{\Lambda} < r_{\mathrm{c}} < \sqrt{3/\Lambda}$.

\begin{figure}[htp]
    \centering
    \includegraphics[width=.99\textwidth]{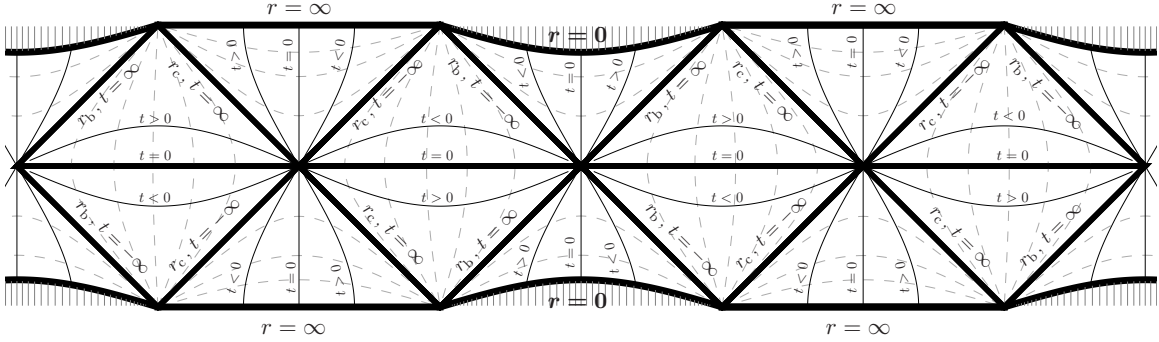}
    \caption{{\small The Kottler-Schwarzschild-de Sitter spacetime KSdS (which extends toward the left and the right ad inifinitum).
        The black holes/white holes $0<r<r_{\mathrm{b}}$ are separated from the future/past cosmological regions 
        $r_{\mathrm{c}} < r < \infty$ by the static regions $r_{\mathrm{b}} < r < r_{\mathrm{c}}$. 
        Solid lines represent hypersurfaces $t=\mathrm{const}$, dashed lines are hypersurfaces $r=\mathrm{const}$.
        Note that the two charts that yield the conformal compactification 
        of the two regions $0<r<r_{\mathrm{c}}$ and $r_{\mathrm{b}}<r<\infty$
        do not agree in the static domains $r_{\mathrm{b}} < r < r_{\mathrm{c}}$; hence, e.g., a dashed line 
        represents $r = \mathrm{const}$ in one chart, but $r= \mathrm{const}^\prime$ in the other.}}
        \label{KSdSfig}
\end{figure}

The static region of the spacetime has an analytic extension reminiscent of the Kruskal extension
of the Schwarzschild spacetime and the de Sitter spacetime. 
On the one hand, the metric is extended beyond the `black hole horizon' $r = r_{\mathrm{b}}$
to arbitrarily small values of $r$, where $r=0$ represents a curvature singularity,
On the other hand, the metric is extended beyond the `cosmological horizon' $r= r_{\mathrm{c}}$
to arbitrarily large values of $r$.
The extended spacetime is commonly depicted using
two charts which cover the regions $0<r<r_{\mathrm{c}}$ and $r_{\mathrm{b}}<r<\infty$, respectively;
for the conformal compactification of these charts see, e.g.,~\cite{Beig/Heinzle:2005}.
In the conformally compactified picture, $r = \infty$ represents conformal infinity (which is spacelike).
As is well-known, the constructed
spacetime corresponding to the union of the two regions $0<r<r_{\mathrm{c}}$ and $r_{\mathrm{b}}<r<\infty$ 
can be smoothly (in fact
analytically) extended in a periodic fashion. 
Thereby one obtains an inextendible,
globally hyperbolic spacetime of topology $\mathbb{R} \times \mathbb{R} \times S^2$, the
Kottler-Schwarzschild-de Sitter spacetime KSdS, see Fig.~\ref{KSdSfig}.

On KSdS there exists an isometric action of $\mathbb{R} \times \mathrm{SO}(3)$. The dashed lines
in Figs.~\ref{KSdSfig} are orbits under the `Killing flow', 
i.e., under the static Killing vector $\xi=\partial_t$. 
The Killing vector $\xi$ is globally defined; it is null on the Killing horizons $r=r_{\mathrm{b}}$ and $r=r_{\mathrm{c}}$
which emanate from the bifurcation 2-spheres at which $\xi$ vanishes. 
The
solid lines in Fig.~\ref{KSdSfig} represent hypersurfaces $t=\mathrm{const}$, which
are totally geodesic as fixed point sets of the discrete isometries (`reflections') $t  + \Delta t \mapsto t - \Delta t$. 
The timelike $t=\mathrm{const}$ hypersurfaces we call $t=\mathrm{const}$ `cylinders'.

\begin{figure}[htp]
    \centering
       \includegraphics[width=0.96\textwidth]{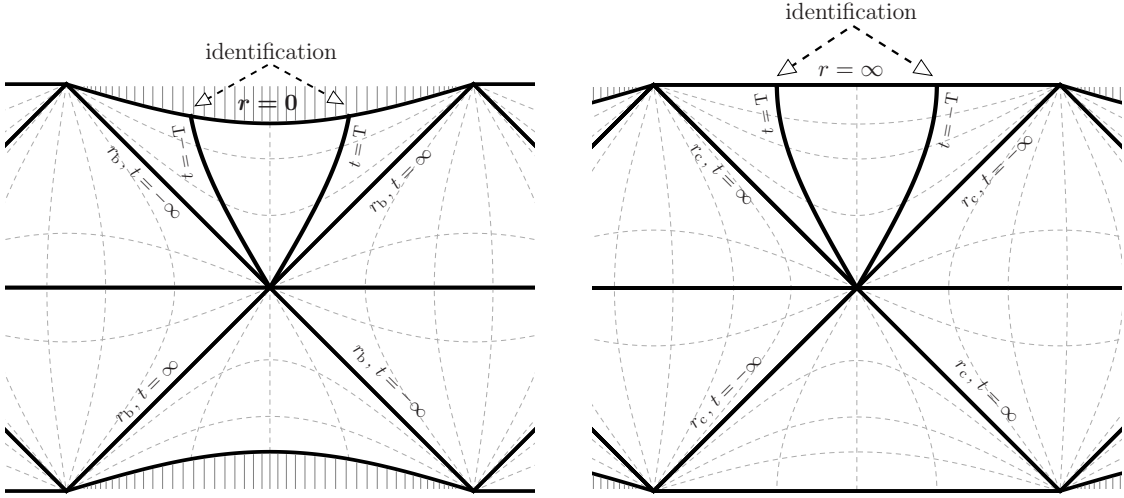}
      \caption{{\small The figure on the l.h.\ side shows the cosmological spacetime BH[T] that arises from the black hole region in the 
          Kottler-Schwarzschild-de Sitter spacetime KSdS.
          Analogously, FC${}_+$[T] is defined by an identification in the future cosmological region, see the r.h.s.\ figure.
          The spacetimes WH[T] and PC${}_+$[T] are not depicted; these are associated with the white hole and the past cosmological region.
          The constructed spacetimes of the Kantowski-Sachs type.}}
        \label{BHFC}
\end{figure}

By a `cosmological spacetime' we mean a globally hyperbolic spacetime
that is spatially compact, i.e., contains a spacelike compact Cauchy hypersurface.
(Then any smooth compact spacelike hypersurface is Cauchy~\cite{Budic:1978}).
Note that we use the term `cosmological spacetime' in a less restrictive manner than~\cite{Bartnik:1988}, 
where the timelike convergence condition (strong energy condition) is assumed in addition.
Of course, this condition is violated in the presence of a positive cosmological constant, i.e.,
\begin{equation}\label{timelikeconvergenceviolated}
R_{\mu\nu} v^\mu v^\nu = -\Lambda < 0 \qquad \text{for all unit timelike vectors $v^\mu$.}
\end{equation}
In the following we will construct a number of classes of `cosmological spacetimes' from KSdS:
 
Consider a \textit{black hole} region $r < r_{\mathrm{b}}$ in KSdS.
By identifying points mapped
to each other by a discrete subgroup of the action under $\xi$, i.e., by identifying 
the cylinders $t=-\mathrm{T}$ and $t= \mathrm{T}$,
we obtain a smooth 
cosmological spacetime of topology
$\mathbb{R} \times S^1 \times S^2$, which we denote by BH[T], see Fig.~\ref{BHFC}.
Similarly, identifying $t=-\mathrm{T}$ and $t= \mathrm{T}$
in a \textit{white hole} region of KSdS, we obtain
a spacetime which we denote by WH[T]
(which is related to BH[T] by time reversal).

Analogously, consider a \textit{future cosmological} region $r_{\mathrm{c}} < r$ in KSdS.
Identifying $t=-\mathrm{T}$ and $t= \mathrm{T}$ we obtain a spacetime
which we call FC${}_+$[T], see Fig.~\ref{BHFC}.
The \textit{past cosmological} region gives rise to PC${}_+$[T]
(which is related to FC${}_+$[T] by time reversal).

The spacetime FC${}_+$[T]/PC${}_+$[T] is future/past geodesically complete, but
geodesically incomplete to the past/future: it possesses
a past/future crushing singularity (in the standard terminology which is due to~\cite{Eardley/Smarr:1979})
corresponding to a past/future mean curvature barrier (in the terminology of, e.g.,~\cite{Gerhardt:2006});
this will become obvious immediately.
BH[T]/WH[T] possesses a past and future crushing singularity, where
the future/past singularity is a curvature singularity.
The constructed spacetimes 
are maximal globally hyperbolic, i.e., 
they agree with the maximal Cauchy development of initial
data defined on any compact spacelike hypersurface.
However, there are non-trivial extensions 
reminiscent of the extensions of the Misner universe (see, e.g.,~\cite{Hawking/Ellis:1973}).

To construct from KSdS a spacetime that is a model for a cosmological spacetime containing a black hole,
we choose $\mathrm{T} \in \mathbb{R}$ and identify
points of equal radius $r$ on a $t=0$ cylinder and a
$t=2 \mathrm{T}$ cylinder in an adjacent copy of the region $0<r < r_{\mathrm{c}}$
on the r.h.\ side. 
Thereby we obtain an inextendible
cosmological spacetime of topology $\mathbb{R}\times S^1\times S^2$, which we call KSdS[T].
KSdS[T] is a smooth, in fact analytic, spacetime, see~\cite[Sec.~1]{Beig/Heinzle:2005} for details.
In KSdS[T] two types of asymptotic behavior coexist: inextendible timelike geodesics either end
in the curvature singularity after finite proper time, or go to scri. 

\begin{figure}[htp]
    \centering
      \includegraphics[width=0.45\textwidth]{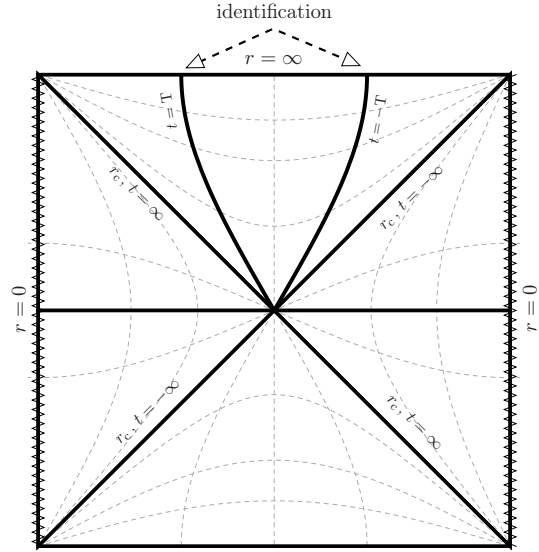}\qquad
      \caption{{\small The Kottler-Schwarzschild-de Sitter spacetime with negative mass $M$. 
          The cosmological spacetime FC${}_-$[T] arises from an identification in the future 
          cosmological region; PC${}_-$[T] is constructed in the past cosmological region.}}
        \label{NegM}
\end{figure}

In the Kottler-Schwarzschild-de Sitter metric~\eqref{KSSdSmetric} 
the mass $M$ need not necessarily be positive (as we have assumed so far).
The Kottler-Schwarzschild-de Sitter spacetime with negative mass, $M < 0$,
has a simpler structure than KSdS:
There is a static region (whose boundary is a naked singularity)
and a past and a future cosmological region $r_{\mathrm{c}} < r < \infty$,
see Fig.~\ref{NegM}.
From this spacetime we can construct cosmological spacetimes:
In the \textit{future cosmological} region $r_{\mathrm{c}} < r$ 
we identify $t=-\mathrm{T}$ and $t= \mathrm{T}$ to obtain
FC${}_-$[T], see Fig.~\ref{NegM}.
The analogous construction in the \textit{past cosmological} region gives rise to PC${}_-$[T].
(The subscripts ${}_\pm$ distinguishing the spacetimes FC${}_\pm$[T] and PC${}_\pm$[T]
thus refer to the positivity/negativity of the mass $M$.)

If $M > 0$ and $9 M^2 \Lambda >1$ in the Kottler-Schwarzschild-de Sitter metric~\eqref{KSSdSmetric}, 
the spacetime does not have a static region and there are no horizons.
We obtain spatially homogeneous cosmological models with topology $\mathbb{R} \times S^1 \times S^2$
which we call SH${}_{\mathrm{e}}$[T] and SH${}_{\mathrm{c}}$[T].

It remains to systematically describe the relationship between the
cosmological spacetimes contructed from 
the Kottler-Schwarzschild-de Sitter metric
and the Kantowski-Sachs spacetimes.
It is immediate that the eight classes of spacetimes
BH[T], WH[T], FC${}_+$[T], FC${}_-$[T], PC${}_+$[T], PC${}_-$[T], SH${}_{\mathrm{e}}$[T], and SH${}_{\mathrm{c}}$[T],
are \textit{Kantowski-Sachs models}; we have 
\begin{equation}
\fourmetric = -|V|^{-1} d r^2 + |V| d t^2 + r^2 \: \metricS \:.
\end{equation}
Since the metric is explicit, it is straightforward to compute
the dynamic quantities $H_D$ and $\Sigma_+$ using~\eqref{HDSig+}.

We find that the spacetimes BH[T] of Fig.~\ref{BHFC} give rise to a one-parameter family of orbits,
which is parametrized by the (positive) mass $M$, that corresponds to
the family of orbits of region BH in Fig.~\ref{KSflow}. 
For small values of $M$ we obtain orbits close to the boundary orbit connecting the Taub point
and the non-flat LRS point. For values of $M$ such that $9 M^2 \Lambda$ is close to $1$
we obtain orbits  that come close to the separatrix orbits and the interior fixed point.
The case is analogous for the spacetimes WH[T], which fill the region WH of Fig.~\ref{KSflow}.

The spacetimes FC${}_+$[T] of Fig.~\ref{BHFC} correspond to the orbits of region FC${}_+$ of Fig.~\ref{KSflow}.
Values of $M$ such that $9 M^2 \Lambda$ is close to $1$ correspond to orbits that come close 
to the separatrix orbits and the interior fixed point. For $M\searrow 0$ 
we obtain orbits in FC${}_+$ close to the special isotropic solution (dashed line). 
The special orbit itself corresponds to a cosmological spacetime, which may be denoted by FC${}_0$[T], 
that is obtained by an identification in the future cosmological region of the de Sitter spacetime.
The family FC${}_-$[T] of Fig.~\ref{NegM} fills the region FC${}_-$ of Fig.~\ref{KSflow}.
For large negative values of $M$ we obtain orbits that are close to the boundary
of the state space. The case of the spacetimes PC${}_\pm$[T] is completely analogous.

The spacetimes SH${}_{\mathrm{e}}$[T] and SH${}_{\mathrm{c}}$[T] correspond to positive values of $M$ with
$9 M^2 \Lambda > 1$. For values of $M$ such that $9 M^2 \Lambda$ is 
only slightly greater than $1$, we obtain orbits in SH${}_{\mathrm{e}}$ and SH${}_{\mathrm{c}}$
that are close to the separatrix orbits and the interior fixed points.
For large values of $M$ we obtain orbits close to the boundary.

Let us make some final comments.
First, we note that the flow of Fig.~\ref{KSflow}
is independent of $\Lambda>0$.
It is possible to interpret $3 M \sqrt{\Lambda}$ as a function
on the state space $(0,1)\times (0,1)$ which is invariant along the orbits.
The net of separatrix orbits (including the two interior fixed points) 
is characterized by $3 M \sqrt{\Lambda} = 1$.
Of course, the separatrix orbits admit an interpretation 
in terms of the Kottler-Schwarzschild-de Sitter metric.
In the case $3 M \sqrt{\Lambda} = 1$ the spacetimes defined from this metric 
are the \textit{Nariai spacetimes}.
We refrain from giving a discussion, but refer to~\cite{Beyer:2009} for a recent study.
(Note that the Kantowski-Sachs models associated with the Nariai spacetimes form a set of measure
zero among all Kantowski-Sachs models.)

Second, a comment in a similar vein.
On the boundaries of the state space of Fig.~\ref{KSflow},
where we have to exclude the fixed points, 
the function $3 M \sqrt{\Lambda}$ takes the values $0$ or $\pm\infty$, respectively.
The solutions on the (`vacuum') boundaries $\Sigma_+ = \pm 1$ (where $3 M \sqrt{\Lambda} = 0$) 
admit a straightforward interpretation: These are the Kantowski-Sachs models
that solve the Einstein vacuum equations with $\Lambda = 0$;
an explicit representation of these models is based on performing an identification
in the black hole/white hole region of the Schwarzschild spacetime.

Third, let us comment on the parameter T, which characterizes the length
of the $S^1$ component of the spacetime. This parameter does not enter
the picture in Fig.~\ref{KSflow}. This is because
it corresponds to the free constant of integration in~\eqref{g11}.

Finally, the statements on past/future crushing singularities 
of the Kantowski-Sachs models BH[T], WH[T], FC${}_+$[T], FC${}_-$[T], PC${}_+$[T], and PC${}_-$[T],
which have been made above,
are now immediate from Fig.~\ref{KSflow} in combination with~\eqref{expans}.

\section{Compact spherically symmetric CMC data}
\label{data}

In this section we briefly discuss the parameter space(s)
of compact spherically symmetric CMC data sets.
For our purposes, a spherically symmetric CMC initial data set
is a triple \mbox{$(I \times S^2, \threemetric_{i j}, k_{i j})$}, where
$I$ is a one-dimensional manifold, $\threemetric_{i j}$ 
a spherically symmetric Riemannian metric,
and $k_{i j}$ a spherically symmetric symmetric tensor with
constant trace, 
subject to the Einstein vacuum constraints with cosmological constant
$\Lambda > 0$.
We find, cf.~\cite{Beig/Heinzle:2005,Beig/OMurchadha:1998},
that these 
data sets are parametrized by $K = \tr k = \mathrm{const}$ (`mean curvature') and a second constant, $C$, through
\begin{subequations}\label{gijKij}
\begin{align}
\label{3-metric}
\threemetric_{i j}\, d x^i d x^j & = d l^2 + r(l)^2 \, \metricS \:, \\
\label{KijinKC}
k_{i j} \, d x^i d x^j & = \frac{K}{3}\: \threemetric_{i j}\, d x^i d x^j + 
C\: \Big( \frac{2}{r(l)^3} \,d l^2 - \frac{1}{r(l)} \,  \metricS \Big) \:,
\end{align}
where $l$ is the coordinate on $I$ and $\metricS$ the standard metric on $S^2$.
(The tensor in brackets in~\eqref{KijinKC} is the standard spherically symmetric transverse traceless tensor.)
The positive function $r(l)$ in~\eqref{3-metric} and~\eqref{KijinKC} is required to be a solution of the differential equation
\begin{equation}\label{rprime2}
r^{\prime\:2} = \Big( \frac{d r}{d l} \Big)^2  = 1 -\frac{2 M}{r} - \frac{\Lambda r^2}{3} + \left( \frac{K r}{3} - \frac{C}{r^2}\right)^2 
=: D(r) \:,
\end{equation}
\end{subequations}
where $M$ is an additional constant. 
The (local) development of the data leads to~\eqref{KSSdSmetric} and $M$ emerges as 
the mass~\cite{Beig/Heinzle:2005}.

\begin{figure}[htp]
  \centering
  \includegraphics[width=\textwidth]{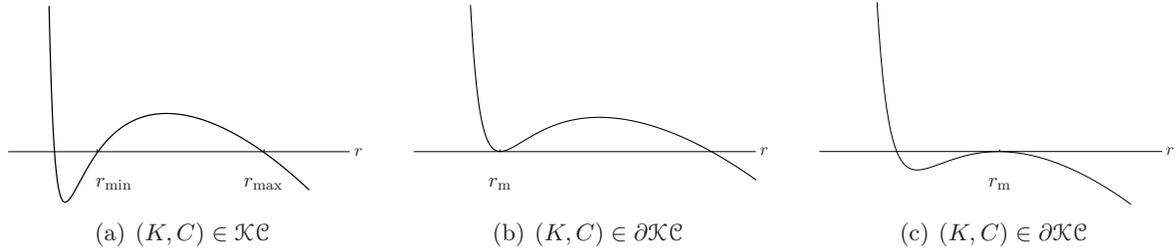}
  \caption{{\small The profiles of $D(r)$ for parameters $(K,C)$ that induce compact CMC data. Subfig.~(a) 
      corresponds to generic CMC data, Subfigs.~(b,c) are borderline cases and generate $r(l) = \rem = \mathrm{const}$
      initial data.}}
  \label{Deltaborderline}
\end{figure}

A spherically symmetric CMC initial data set 
is \textit{compact}, if $I \cong S^1$. 
This is the case if $K$ and $C$ are such that $D(r)$ has two positive (simple) zeros, $r_{\min}$ and $r_{\max}$, 
and $D(r) > 0$ in the interval $(r_{\min}, r_{\max})$; see Fig.~\ref{Deltaborderline}(a).
This leads to a periodic solution $r(l)$ of~\eqref{rprime2}  
that oscillates between $r_{\min}$ and $r_{\max}$;
we denote the period by $2 L$,
\begin{equation}\label{periodL}
L = \int\limits_{r_{\min}}^{r_{\max}} D^{-1/2}(r)\: d r\:.
\end{equation}
W.l.o.g.\ we assume that
$r(0) = r_{\min}$, so that $r(\pm L) = r_{\max}$;
it follows that $r(l)$ is even.
By the natural identification of $l=-L$ and $l=L$, the domain of the 
function $r(l)$ becomes $S^1$ and the
CMC initial data set $(S^1\times S^2, \threemetric_{i j}, k_{i j})$ is compact.

There are `degenerate' cases: If $K$ and $C$ are such that $D(r)$
possesses a root $\rem$ of multiplicity two (or three), see Figs.~\ref{Deltaborderline}(b) and~\ref{Deltaborderline}(c),
then $r(l) \equiv \rem$ is a constant solution of~\eqref{rprime2}.
Clearly, these solutions generate compact CMC initial data sets $(S^1\times S^2, \threemetric_{i j}, k_{i j})$
where the `length' $2 L$ of the $S^1$ component is arbitrary.

\subsection{Data with $\bm{r(l) \equiv \rem =} \mathbf{const}$}
\label{rconstsubsec}

Compact CMC initial data sets 
with $r(l) \equiv \rem = \mathrm{const}$
are obtained from~\eqref{rprime2} as solutions of the algebraic
equations $D(\rem) =0$, $(d D/d r)(\rem) = 0$, cf.~Figs.~\ref{Deltaborderline}(b) and~\ref{Deltaborderline}(c).
The parameters $K$ and $C$ associated with these data sets are 
\begin{equation}\label{KCrconst}
\big(K, C\big) = \big(\Kem(\rem),\Cem(\rem)\big) = \pm \left(\frac{1}{\rem^2} \frac{3 M - \rem}{\sqrt{-V(\rem)}} 
- \frac{3}{\rem}\,\sqrt{-V(\rem)}\,,\;
\frac{\rem}{3}\,\frac{3 M - \rem}{\sqrt{-V(\rem)}}\: \right) \:,
\end{equation}
where $V(r)$ is the function of~\eqref{KSSdSmetric}.

The admissible range of $\rem$ 
depends on the values of $\Lambda$ ($>0$) and $M$ which determine $V(r)$.
In the following we concentrate on the case $M > 0$, $9 M^2 \Lambda < 1$ (which is 
the most intricate one).
Accordingly, the admissible range of $\rem$  is $(0, r_{\mathrm{b}}) \cup (r_{\mathrm{c}},\infty)$.

\begin{remark}
The CMC data sets with $r(l) \equiv \rem = \mathrm{const}$
are associated with the spacelike hypersurfaces $r \equiv \rem$ in KSdS,
$\rem\in (0, r_{\mathrm{b}}) \cup (r_{\mathrm{c}},\infty)$.
The upper [lower] sign in~\eqref{KCrconst} refers to hypersurfaces in the black [white] hole 
and the past [future] cosmological region of KSdS.
\end{remark}

\begin{figure}[htp]
         \centering	  
	\includegraphics[width=0.85\textwidth]{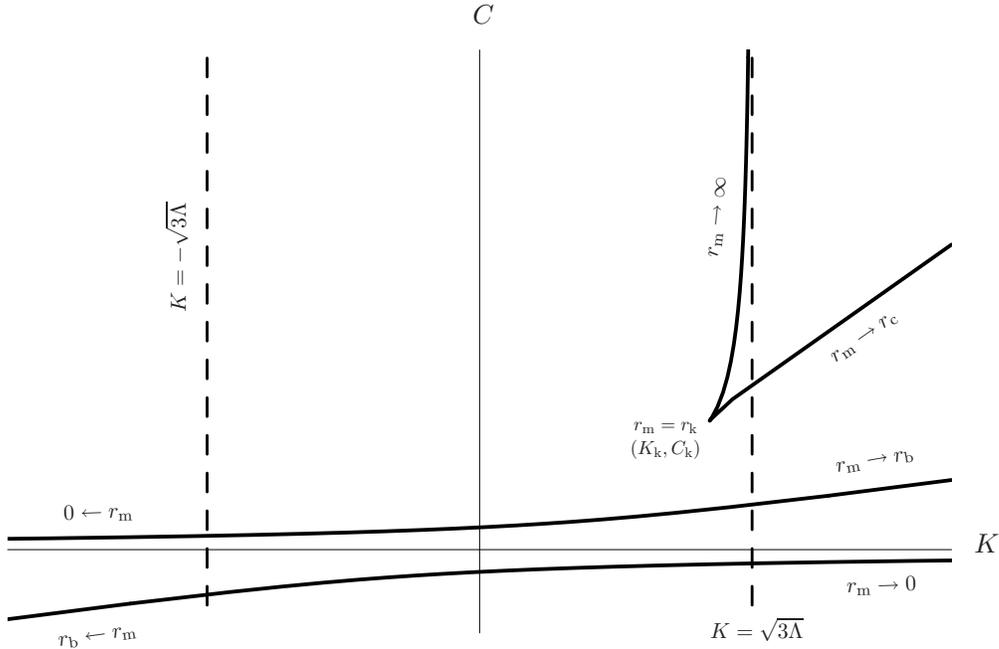}	
	\caption{{\small The curves representing the $r(l)=\rem = \mathrm{const}$ CMC initial data sets 
            (in the case \mbox{$M > 0$}, \mbox{$9 M^2 \Lambda = 81/100$}).
            The mirror image of the `wedge' in the upper right corner, which is obtained by reflection at the origin, is not depicted.}}
    	\label{Cbtdu}
\end{figure}

Equation~\eqref{KCrconst} describes two pairs of curves, $\rem \mapsto \big(\Kem(\rem),\Cem(\rem)\big)$, one pair
corresponding to $\rem \in (0, r_{\mathrm{b}})$ and another pair with $\rem \in (r_{\mathrm{c}},\infty)$;
these curves are depicted in Fig.~\ref{Cbtdu}. (It is straightforward to prove
the monotonicity properties and the asymptotic behavior of these curves suggested in this figure.)
A conspicuous feature is the cusp of the curve \mbox{$(r_{\mathrm{c}}, \infty) \ni \rem \mapsto \big(\Kem(\rem)>0,\Cem(\rem)>0\big)$}, 
which means that there exists a unique $
\rk > r_{\mathrm{c}}$ such that
\mbox{$d \Kem/ d\rem(\rk) = 0$} and \mbox{$d \Cem/ d\rem(\rk) = 0$}; the value of $\rk$ is determined
by the equation
\begin{equation}\label{rkdef}
\rk^2 \:V(\rk) + ( 3 M - \rk )^2 = 0\:.
\end{equation}
The value of $|K|$ at $\rk$ is the smallest possible value of the (modulus of the) mean curvature
within the class $\rem \in (r_{\mathrm{c}},\infty)$; the same is true for the value of $|C|$;
we denote these by $K_{\mathrm{k}}$ and $C_{\mathrm{k}}$, respectively.
It is not difficult to show that $\sqrt{\Lambda} < K_{\mathrm{k}} < \sqrt{3 \Lambda}$.

\begin{remark}
In terms of the function $D(r)$, $\rk$ is determined by the conditions
$D(\rk) =0$, $(d D/d r)(\rk) = 0$, and
$(d^2 D/d r^2)(\rk) = 0$.
Hence, for $\pm(K_{\mathrm{k}}, C_{\mathrm{k}})$, and only for these values,
the function $D(r)$ has a zero of third order, which corresponds to
the borderline case between Figs.~\ref{Deltaborderline}(b) and~\ref{Deltaborderline}(c).
\end{remark}

\subsection{The parameter space of compact CMC data}
\label{Sec:KCi}

We define the parameter space $\pmb{\mathscr{KC}}$ of compact CMC data
to be the interior of the set of pairs $(K,C)$ that generate compact CMC data.
For $(K,C)$ to be in $\mathscr{KC}$, $D(r)$  
must have the form depicted in Fig.~\ref{Deltaborderline}(a);
then $r(l)$ is an even function that oscillates between $r_{\min}$ and $r_{\max}$,
where the period $2 L$ is given by~\eqref{periodL}.

The parameter space $\mathscr{KC}$ is the disjoint union
of three open connected domains, 
\begin{equation}\label{KCall}
\mathscr{KC} = \mathscr{KC}_0 \cup \mathscr{KC}_1 \cup \mathscr{KC}_{-1}\:,
\end{equation}
which are defined by the relations
\begin{subequations}\label{KCdef}
\begin{alignat}{2}
\label{KC0def}
& (K,C) \in \mathscr{KC}_{0} \: & & \Longleftrightarrow\:
r_{\min} \leq r_{\mathrm{b}}< r_{\mathrm{c}} \leq r_{\max} \:,\\[1ex]
\label{KC1def} 
& (K,C)\in \mathscr{KC}_{\pm 1} \: & & \Longleftrightarrow\:
\,r_{\mathrm{c}} < r_{\min} < r_{\max} \:.
\end{alignat}
\end{subequations}
The domain $\mathscr{KC}_{-1}$ arises from $\mathscr{KC}_{1}$ by the inversion at the origin,
while $\mathscr{KC}_0$ is invariant under the inversion.

Let us elaborate on~\eqref{KCdef}.
Since the function $D(r)$ is non-negative in $(r_{\mathrm{b}}, r_{\mathrm{c}})$, see~\eqref{rprime2},
the roots $r_{\min}$ and $r_{\max}$ either satisfy $r_{\min} \leq r_{\mathrm{b}}< r_{\mathrm{c}} \leq r_{\max}$, cf.~\eqref{KC0def}, 
or
$r_{\mathrm{c}} \leq r_{\min} < r_{\max}$, or $r_{\min} < r_{\max} \leq r_{\mathrm{b}}$.
To obtain~\eqref{KCdef} we need to show that the subcase $r_{\mathrm{c}} = r_{\min} < r_{\max}$
and the entire third scenario $r_{\min} < r_{\max} \leq r_{\mathrm{b}}$ are excluded:
First, if we assume that $r_{\min} = r_{\mathrm{c}}$, then
$0= D(r_{\min}) = (K r_{\mathrm{c}}/3 - C/r_{\mathrm{c}}^2)^2$, and hence
$(d D/d r)(r_{\min}) = (d V/d r)(r_{\mathrm{c}}) < 0 $; however, this contradicts
the definition of $r_{\min}$, which includes the condition $(dD/d r)(r_{\min}) >0$.
Second, we note that $D(r)$ can possess at most three zeros,
of which $r_{\max}$ is the largest. This follows
from an analysis of the critical points of $D(r)$, see~\cite[Appendix~A]{Beig/Heinzle:2005}.
Accordingly, $D(r)$ must be negative for all $r>r_{\max}$.
This being the case, the assumption $r_{\min} < r_{\max} \leq r_{\mathrm{b}}$ ($< r_{\mathrm{c}}$)
immediately leads to a contradiction, since 
$D(r_{\mathrm{c}}) \geq 0$. This establishes the implicit claims of~\eqref{KCdef}.

\begin{figure}[htp]
  \centering
        \includegraphics[width=0.9\textwidth]{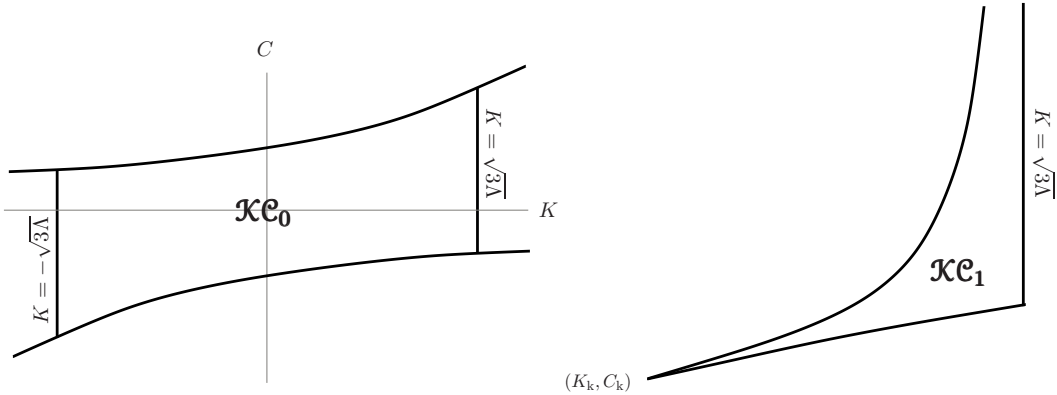}
	\caption{{\small The parameter spaces $\mathscr{KC}_0$ and $\mathscr{KC}_{1}$ of compact CMC data are bounded by
            the curves of Fig.~\ref{Cbtdu} that represent $r(l) = \mathrm{const}$ initial data and
            by the straight line(s) $K = \pm \sqrt{3 \Lambda}$.}}
    	\label{KC01}
\end{figure}

The parameter spaces $\mathscr{KC}_0$ and $\mathscr{KC}_1$ are depicted in Fig.~\ref{KC01}.
Among the boundaries of these spaces 
are the curves of Fig.~\ref{Cbtdu} that represent the pairs $(K,C)$
that generate $r(l) = \rem = \mathrm{const}$ initial data.
For these `critical' values of $(K,C)$ the 
profile of the function $D(r)$ is a borderline case, see Fig.~\ref{Deltaborderline}.
There are arbitrarily small perturbations of Fig.~\ref{Deltaborderline}(b) and~\ref{Deltaborderline}(c) that lead to the profile~\ref{Deltaborderline}(a)
and thus to compact CMC data, but there are other perturbations that lead
to a profile of $D(r)$ that is not associated with compact CMC data.
The remaining boundary components of $\mathscr{KC}_0$ and $\mathscr{KC}_1$
are the straight lines $K = \pm\sqrt{3 \Lambda}$.
For \mbox{$\mathscr{KC} \ni (K,C) \rightarrow (\pm\sqrt{3\Lambda}, C)$}, we find $r_{\max} \rightarrow \infty$, 
cf.~Fig.~\ref{Deltaborderline}(a), because
the function $D(r)$ is positive for large $r$, if $|K| \geq \sqrt{3 \Lambda}$.

\begin{remark}
The analysis of~\eqref{KCrconst} and the parameter space of compact CMC data 
is simpler in the remaining cases. In the case $M < 0$ we find
$\mathscr{KC} = \mathscr{KC}_1 \cup \mathscr{KC}_{-1}$, where
the space $\mathscr{KC}_1$ is a `wedge' whose properties are analogous to
those sketched in Fig.~\ref{KC01}.
The case $M > 0$, $9 M^2 \Lambda > 1$ is similar; however, we refrain 
from going into the details.
\end{remark}

\section{Embeddings}
\label{embeddings}

In this section we investigate whether 
compact CMC initial data sets are embeddable
in the Kantowski-Sachs spacetimes and/or the cosmological Kottler-Schwarzschild-de Sitter spacetime KSdS[T].

Consider a compact CMC initial data set 
$(S^1 \times S^2 , \threemetric_{i j}, k_{i j})$ represented by $\Lambda$, $M$, and a pair \mbox{$(K,C) \in \mathscr{KC}$}.
The universal covering $(\mathbb{R} \times S^2 , \threemetric_{i j}, k_{i j})$ of the data 
is embeddable as a CMC hypersurface $\tilde{\mathcal{S}}$ in KSdS through
\begin{equation}\label{tofl}
r = r(l) \:,\qquad t = t(l) := \int\limits_0^l V^{-1}(r(\hat{l})) \:
\left(\frac{K r(\hat{l})}{3} -\frac{C}{r(\hat{l})^2}\right)  d \hat{l} \:,
\end{equation}
see~\cite{Beig/Heinzle:2005}; 
the integral is understood in the principal value sense.
The embedding~\eqref{tofl} is not unique: $r=r(l)$, $t = t(l) +\mathrm{const}$ defines
a one-parameter family of embeddings of the data; this reflects 
the continuous isometry of the spacetime generated by the Killing vector $\partial_t$.

\begin{remark}
The hypersurface $\tilde{\mathcal{S}}$ represented by~\eqref{tofl} 
is reflection symmetric, i.e., invariant under the discrete isometry 
$t\mapsto -t$ of the spacetime.
The requirement of reflection symmetry is a convenient way of fixing a
representative within the one-parameter family of embeddings $r=r(l)$, $t = t(l) +\mathrm{const}$:
$\tilde{\mathcal{S}}$ is the unique reflection symmetric embedding of the data. 
\end{remark}

The (future pointing) unit normal of $\tilde{\mathcal{S}}$ in KSdS is given by
\begin{equation}\label{unitnor}
n^\mu \partial_\mu = r^\prime \,V^{-1}\, \frac{\partial}{\partial t} + 
\,\left(\frac{K r}{3} -\frac{C}{r^2}\right)\:\frac{\partial}{\partial r} \:,
\end{equation}
where the prime denotes the derivative w.r.t.\ $l$.
It is straightforward to check that $\nabla_\mu n^\mu = K$.
The Killing vector $\xi = \partial_t$ of KSdS admits a 
decomposition into a lapse function $\alpha_\xi$ and a shift vector $X_\xi^i \partial_i$:
From~\eqref{tofl} and~\eqref{unitnor} we obtain
$\alpha_\xi = r^\prime(l)$ and 
$X_\xi^i \partial_i  =-(Kr/3-C/r^2) \partial_l$.
We now make a fundamental definition.

\begin{definition}
Let $(K,C) \in \mathscr{KC}$ and $L$ be the (half-)period~\eqref{periodL}
characterizing the compact CMC initial data set.
We define
\begin{equation}\label{Tdef}
\mathcal{T} := |t(L)| = 
\Big|\int\limits_0^L V^{-1}(r(l)) \:\left(\frac{K r(l)}{3} -\frac{C}{r(l)^2}\right)  d l \Big|
= \Big|\hspace{-1em}\int\limits_{\mathrm{range}(r(l))} \hspace{-1em} V^{-1}(r) D^{-1/2}(r)
\:\left(\frac{K r}{3} -\frac{C}{r^2}\right)  d r\Big|\:.
\end{equation}
\end{definition}

The value of the quantity $\mathcal{T}$ is a property of the initial data set.
It is possible (and useful) to regard $\mathcal{T}$ as a function of $(K,C)$ on $\mathscr{KC}$.

In~\cite[Sec.~3]{Beig/Heinzle:2005} the following statement has been proved:
A compact CMC initial data set \mbox{$(S^1 \times S^2, \threemetric_{i j}, k_{i j})$} associated with a pair $(K,C) \in \mathscr{KC}_0$
is embeddable as a smooth CMC hypersurface $\mathcal{S}$ in KSdS[T] 
if and only if $\mathrm{T} = \mathcal{T}$. The hypersurface $\mathcal{S}$ is then a Cauchy hypersurface and the 
embedding is unique modulo the Killing flow.
We now prove an analogous statement for $(K,C) \in \mathscr{KC}_{\pm 1}$.

\begin{proposition}\label{KC1embedprop}
Let $M < 0$, or $M > 0$ with $9 M^2 \Lambda < 1$.
Consider a compact CMC initial data set $(S^1 \times S^2, \threemetric_{i j}, k_{i j})$
associated with $(K,C) \in \mathscr{KC}_1$. 
This data set is embeddable as a smooth CMC hypersurface $\mathcal{S}$ ($=\mathcal{S}_1$) in the 
Kantowski-Sachs spacetime FC${}_-$[T] or FC${}_+$[T], depending
on the sign of $M$, if and only if\/ $T = \mathcal{T}$.
The $n$-fold covering of the data is embeddable 
as a smooth CMC hypersurface $\mathcal{S}_n$ in FC${}_\pm$[T]
if and only if\/ $T = n \mathcal{T}$. 
The embedded hypersurfaces are Cauchy hypersurfaces,
and the embeddings are unique modulo the Killing flow.
\end{proposition}

\begin{figure}[htp]
	\centering
	\includegraphics[width=0.9\textwidth]{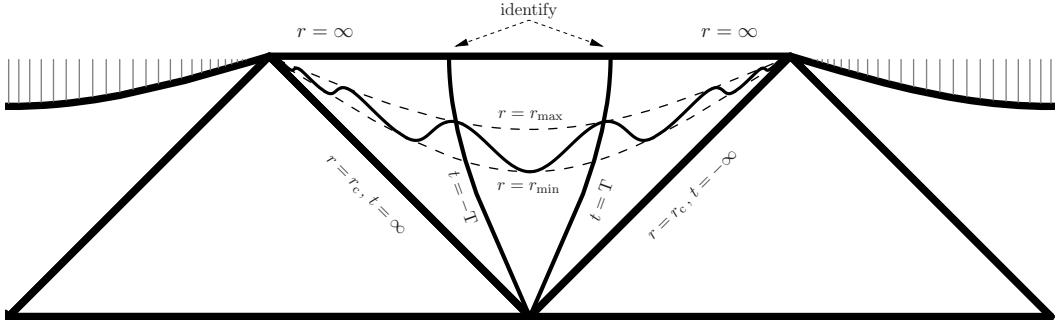}	  
	\caption{{\small A CMC hypersurface associated with $(K,C) \in \mathscr{KC}_1$ (and $M >0$, $9 M^2 \Lambda < 1$)
	    appears as a ``corrugated sheet'' confined between the 
	    $r=\mathrm{const}$ hypersurfaces $r=r_{\min}$ and $r= r_{\max}$. In FC${}_+$[T] with $\mathrm{T} = \mathcal{T}$, it 
	    is a smooth Cauchy hypersurface.}}
    	\label{corrugatedsheet}
\end{figure}

\proof
Consider the universal covering of the data and the embedded
hypersurface $\tilde{\mathcal{S}}$ in KSdS as given by~\eqref{tofl}. 
Since $r(l)$ oscillates between $r_{\min}$ and $r_{\max}$ 
where $r_{\mathrm{c}} < r_{\min} < r_{\max}$, see~\eqref{KC1def},
$\tilde{\mathcal{S}}$ is entirely contained in a cosmological region of KSdS.
The assumption $(K,C) \in \mathscr{KC}_1$ implies $K r/3-C/r^2>0$ and we infer from~\eqref{unitnor} that
$\tilde{\mathcal{S}}$ is contained in a future cosmological region of KSdS, see~Fig.~\ref{corrugatedsheet}. 
The hypersurface $\tilde{\mathcal{S}}$ is invariant under
the reflections $m\mathcal{T} + t \mapsto m \mathcal{T} -t$ for all $m\in\mathbb{Z}$;
this is because $r(l)$ is even about $l=m L$ $\forall m$. 
Since the unit normal vectors of $\tilde{\mathcal{S}}$ at $l= m L$ must
be invariant under these reflections as well, they must be tangential
to the fixed point sets $t= m \mathcal{T}$ of these discrete isometries. We conclude
that $n^\mu \partial_\mu \propto \partial_t$ at $l=m L$. Therefore,
the identification of the cylinders $t={-n} \mathcal{T}$ and $t = n \mathcal{T}$ ($n\in \mathbb{N}$) in KSdS
turns $\tilde{\mathcal{S}}$ into a smooth CMC hypersurface $\mathcal{S}_n$ 
in FC${}_\pm$[$n \mathcal{T}$] (where the sign is determined by the sign of $M$). 
Hence, since $\mathcal{S}_n$ is isometric to the $n$-fold covering of the initial data, we have found that
the $n$-fold covering of the data is embeddable as a smooth CMC hypersurface $\mathcal{S}_n$ 
in FC${}_\pm$[T], $\mathrm{T} = n \mathcal{T}$.
In spacetimes FC${}_\pm$[T] with $\mathrm{T} \neq  n \mathcal{T}$ $\forall n$,  however, 
the identification of $t=-\mathrm{T}$ with $t = \mathrm{T}$ 
is inconsistent with the identifications of $l=-n L$ with $l=n L$ 
on the level of the data, which is an obstruction to (smooth) embeddability.
The proof of the remaining claims is trivial.
\proofend

\begin{remark}
The statement of the proposition holds analogously for $(K,C) \in \mathscr{KC}_{-1}$ 
if we replace FC${}_\pm$[T] (future cosmological) by PC${}_\pm$[T] (past cosmological).
\end{remark}

\begin{remark}
The case $M = 0$ is completely analogous as well. The ($n$-fold covering of the) data is embeddable 
in the cosmological spacetime FC${}_0$[T] obtained from the de Sitter spacetime, cf.~section~\ref{KSdSsubsec}, 
if $\mathrm{T} = n \mathcal{T}$.
We do not consider the case $M > 0$, $9 M^2 \Lambda > 1$; however, we expect a result that is analogous
to Proposition~\ref{KC1embedprop}, where FC${}_\pm$[T] and PC${}_\pm$[T] are replaced by SH${}_{\mathrm{e}}$[T]
and SH${}_{\mathrm{c}}$[T].
\end{remark}

\section{Slicings}
\label{Sec:CdST}

The results of section~\ref{embeddings} leave open
the question which of the Kantowski-Sachs spacetimes FC${}_\pm$[T] and PC${}_\pm$[T]
actually contain CMC hypersurfaces.
Another issue that will be discussed in this section 
is the question of whether the embedded CMC hypersurfaces
evolve to form slicings or foliations.
The interesting case is that of the Kantowski-Sachs spacetimes FC${}_\pm$[T] and PC${}_\pm$[T].

Let us begin, however, by investigating the Kantowski-Sachs spacetimes of
the classes BH[T] and WH[T].
The results of sections~\ref{data} and~\ref{embeddings} show that
there do not exist (spherically symmetric) CMC hypersurfaces
in these spacetimes other than the $r = \mathrm{const}$ hypersurfaces;
see~\eqref{KCall} and~\eqref{KCdef} in particular.
The $r = \mathrm{const}$ hypersurfaces are associated with the $r(l) = \rem = \mathrm{const}$
initial data sets which are represented by pairs $(K,C)$
determined by~\eqref{KCrconst} with $0 < \rem < r_{\mathrm{b}}$, see Fig.~\ref{Cbtdu}.
The $r = \mathrm{const}$ hypersurfaces form a global foliation
of BH[T]/WH[T], and the mean curvature is a monotone function along
this foliation, which we infer from Fig.~\ref{KSflow} in combination 
with~\eqref{expans} or from Fig.~\ref{Cbtdu}.
In other words, \textit{the Kantowski-Sachs spacetimes {\bfseries BH[T]} and {\bfseries WH[T]} possess 
a global CMC time function} and thus a unique global CMC foliation.

The Kantowski-Sachs spacetimes of the classes FC${}_\pm$[T] and PC${}_\pm$[T]
exhibit a richer structure of CMC slicings.
Let us concentrate on the spacetimes FC${}_\pm$[T].
The $r= \rem = \mathrm{const}$ hypersurfaces, $r_{\mathrm{c}} < \rem < \infty$,
form a global CMC foliation of FC${}_\pm$[T]. 
This foliation is reflection symmetric and invariant under the Killing flow.
However, this global CMC foliation does not define a CMC time function.
Note that the mean curvature is not monotone and attains a minimal value
$\Kk$ along the foliation, where $\Kk = K(\rk)$ 
is given by~\eqref{KCrconst} and~\eqref{rkdef};
it satisfies $\sqrt{\Lambda} < \Kk < \sqrt{3 \Lambda}$.
More specifically,
from Fig.~\ref{KSflow} in combination 
with~\eqref{expans} or Fig.~\ref{Cbtdu} we see that
$K$ ($=\tr k$) decreases monotonically from $\infty$ to $\Kk$ as $\rem$ increases from $r_{\mathrm{c}}$ to $\rk$;
at $\rem = \rk$ it attains its minimal value $\Kk$;
as $\rem$ increases further, $K$ is monotonically increasing as well and converges to $\sqrt{3 \Lambda}$
in the limit $\rem \rightarrow \infty$.
In some of the Kantowski-Sachs spacetimes FC${}_\pm$[T],
this global CMC foliation is the unique CMC foliation:

\begin{theorem}\label{CdSThm1}
Consider a Kantowski-Sachs spacetime FC${}_\pm$[T] or PC${}_\pm$[T] with
\[
\mathrm{T} \leq \sqrt{\frac{3}{\Lambda}}\;\pi \:. 
\]
Then the unique (spherically symmetric) CMC foliation is the global foliation 
of hypersurfaces \mbox{$r= \rem = \mathrm{const}$}, where $r_{\mathrm{c}} < \rem < \infty$.
\end{theorem}

Recall that $\mathrm{T}$ is a measure of the length of the $S^1$ factor
of the Kantowski-Sachs spacetime.
The situation is more intricate and more interesting for Kantowski-Sachs
spacetimes with larger $\mathrm{T}$.

\begin{theorem}\label{CdSThm}
Consider a Kantowski-Sachs spacetime FC${}_\pm$[T] or PC${}_\pm$[T] with
\[
\mathrm{T} > \sqrt{\frac{3}{\Lambda}}\;\pi \:,
\]
and let $n^\star\in \mathbb{N}$ be the greatest integer strictly 
less than $\mathrm{T} \,\pi^{-1} \sqrt{\Lambda/3}$. 
Then the spacetime contains the global CMC foliation 
of $r= \rem = \mathrm{const}$ hypersurfaces and, in addition, 
$n^\star$ distinct families of (maximally extended) CMC slicings.
\end{theorem}

\begin{remark}
The existence of `families' of CMC slicing 
is due to the fact that the embedded CMC hypersurfaces are not unique.
The one-parameter family of isometries associated with the Killing vector $\xi = \partial_t$
turns each CMC hypersurfaces into a one-parameter family of CMC hypersurfaces
(where of course each hypersurface of the family is characterized by the same parameters $K$ and $C$).
Requiring reflection symmetry
about the $t = 0$ cylinder is a convenient way of fixing 
a unique representative of the family, see section~\ref{embeddings}.
Therefore, Theorem~\ref{CdSThm} can be restated:
Under the given assumptions, 
\textit{FC${}_\pm$[T] (or PC${}_\pm$[T])
contains the global CMC foliation 
of $r= \rem = \mathrm{const}$ hypersurfaces and
$n^\star$ slicings of reflection symmetric compact CMC hypersurfaces}.
The remaining compact CMC slicings
arise from these reflection symmetric slicings by
an appropriate admixture of the Killing flow.
\end{remark}

\begin{remark}
In addition to the statement of Theorem~\ref{CdSThm} we will see that 
along the nontrivial slicings the mean curvature is a strictly monotone 
function where $K$ ranges over a subinterval of $(\Kk, \sqrt{3\Lambda} )$.
\end{remark}

The remainder of this section is concerned with the \textit{proof} of
the theorems and the additional claims.
The first part of the proof is essentially identical to 
the proof of Theorem~4.1 of~\cite{Beig/Heinzle:2005}. 
Let us thus merely sketch those arguments:
The starting point is Proposition~\ref{KC1embedprop}, which 
establishes the existence of (reflection symmetric) compact CMC hypersurfaces $\mathcal{S}_n$ in (certain) FC${}_\pm$[T] spacetimes.
The proof that each of these hypersurfaces evolves into a unique (reflection symmetric)
local slicing $\mathcal{S}_n(\tau)$, where $\tau$ is a parameter of the slicing such that $\mathcal{S}_n(0) =  \mathcal{S}_n$,
is based on an analysis of the lapse equation, see, e.g.,~\cite{Choquet-Bruhat/Fischer/Marsden:1979,Beig/Heinzle:2005},
\begin{equation}\label{lapse}
\Delta_g \alpha + a \alpha = \dot{K}  \quad\text{with}\quad
a = \Lambda - k_{i j} k^{i j} = \Lambda -\frac{K^2}{3} - \frac{6 C^2}{r^6}\:,
\end{equation}
where $\Delta_g$ is the Laplacian of the induced metric~\eqref{3-metric}
and $\dot{K} = \dot{K}(\tau)$;
the $\tau$-dependencies are suppressed.
In~\cite[Thm.\ 4.1]{Beig/Heinzle:2005} it is shown that~\eqref{lapse}
possesses a unique even solution $\alpha: S^1\rightarrow \mathbb{R}$ provided that $\dot{K}\neq 0$,
which implies the existence (and uniqueness) of (reflection symmetric) local CMC slicings.

If the hypersurface $\mathcal{S}_n$ corresponds to the point $(K, C)$ in the parameter space $\mathscr{KC}_1$,
see Fig.~\ref{KC01},
then the constructed local slicing $\mathcal{S}_n(\tau)$ corresponds to a (piece of a) smooth curve in 
$\mathscr{KC}_1$, which passes through that point.
Smoothness follows
from smoothness of $k_{i j}$ and $K = \tr k$ along $\mathcal{S}_n(\tau)$, because
\begin{equation}\label{KijKijtau}
k_{i j}(\tau;r) k^{i j}(\tau;r) - \frac{K^2(\tau)}{3}  = \frac{6 C^2(\tau)}{r^6}\:.
\end{equation}

To determine these curves in $\mathscr{KC}_1$ which represent CMC slicings,
we differentiate~\eqref{KijKijtau} and use the evolution and the lapse equation on the l.h.s.\
to obtain
\begin{equation}\label{Cdoteq}
r^\prime \alpha^\prime - r^{\prime\prime} \alpha =
\frac{\dot{K} r}{3} -\frac{\dot{C}}{r^2}\:.
\end{equation}
In this equation, $\alpha$ is the (unique) even solution of~\eqref{lapse} and 
the prime denotes the derivative w.r.t.\ $l$.
Evaluation of~\eqref{Cdoteq} at $l = 0$, which corresponds to $r = r_{\min}$,
and use of~\eqref{rprime2} yields
\begin{equation}\label{Cdotonemin}
\frac{\dot{C}}{r^2_{\min}} = 
\frac{1}{2} \: \alpha(r_{\min})\: \frac{d D}{d r}(r_{\min}) + \frac{\dot{K} r_{\min}}{3} \,,
\end{equation}
see~\cite{Beig/Heinzle:2005} for details.
We infer that the tangent vectors of the curves $(K,C)(\tau)$ in $\mathscr{KC}_1$ which represent CMC slicings
are determined by~\eqref{Cdotonemin}.

\begin{figure}[htp]
        \centering
	\includegraphics[width=0.8\textwidth]{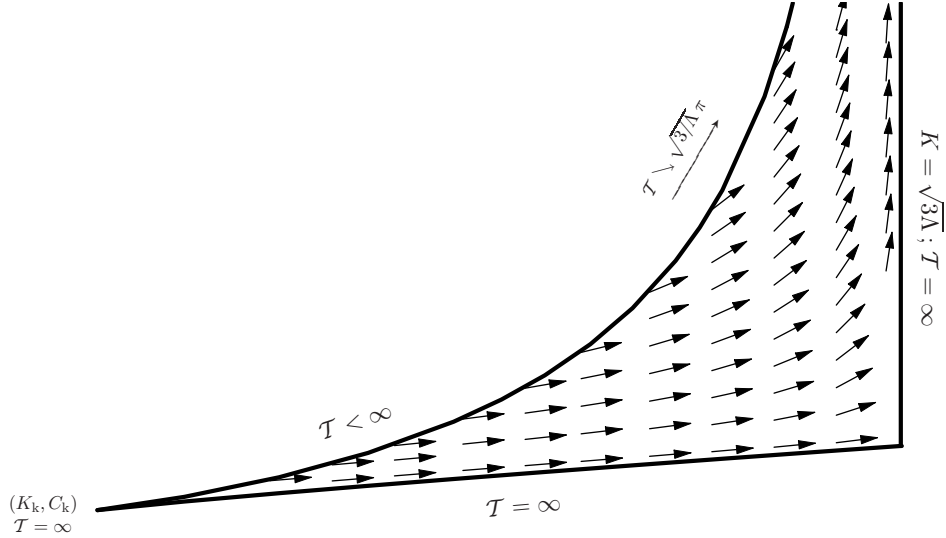}
	\caption{{\small The oriented direction field~\eqref{Cdotonemin} on $\mathscr{KC}_1$ for the case $\Lambda=1$, $M=1/4$.
            The integral curves of this direction field are equipotential lines of the function $\mathcal{T}$ on $\mathscr{KC}_1$
            and represent compact CMC slicings in Kantowski-Sachs spacetimes FC${}_\pm$[T], $\mathrm{T} = n \mathcal{T}$, $n\in\mathbb{N}$.}}
    	\label{KC1flowN}
\end{figure}

Summarizing, compact CMC slicings are represented in $\mathscr{KC}_1$ by
the integral curves of
the `oriented direction field' $(\dot{K}>0, \dot{C})$ given by~\eqref{Cdotonemin}, see Fig.~\ref{KC1flowN}.
Maximal extension of the integral curves of the oriented direction
field on $\mathscr{KC}_1$ corresponds to maximal extension of the CMC slicing in the spacetime.

Consider an integral curve $(K,C)(\tau)$ of the oriented direction field~\eqref{Cdotonemin}.
By construction, the quantity $\mathcal{T}$ of~\eqref{Tdef} is constant along this curve and
the associated CMC slicing $\mathcal{S}(\tau)$ is a slicing in FC${}_\pm$[$\mathcal{T}$].
However, every point $(K,C) \in \mathscr{KC}_1$ represents the original CMC data set
as well as its (finite) coverings. Therefore, the integral curve represents
not only a CMC slicing  $\mathcal{S}(\tau)$ ($=\mathcal{S}_1(\tau)$) in FC${}_\pm$[$\mathcal{T}$], but
also a CMC slicing $\mathcal{S}_n(\tau)$ in FC${}_\pm$[$n \mathcal{T}$] for all $n\in\mathbb{N}$;
each hypersurface of the slicing $\mathcal{S}_n(\tau)$ corresponds
to the $n$-fold covering of the original data.

To establish the claims of the theorems (and to obtain additional
information on the properties of the CMC slicings)
it is necessary to understand 
the behavior of $\mathcal{T}$, viewed as a function of $(K,C) \in \mathscr{KC}_1$,
see~\eqref{Tdef}.

First, we investigate the behavior of $\mathcal{T}$ on the boundary; 
the relevant statements are proved in appendix~\ref{Tonboundary}:
Consider a curve in $\mathscr{KC}_1$ converging to a point on the boundary,
which is either of following: A point on the right boundary with $K = \sqrt{3 \Lambda}$;
a point on the lower boundary (including the cusp), which is uniquely represented by a value of $\rem$ in
an interval $(r_\ast, \rk]$ (where $r_\ast$ is known explicitly, see appendix~\ref{Tonboundary}); or a point on the left boundary 
which is uniquely represented by a value of $\rem$ in
the interval $(\rk,\infty)$; see Fig.~\ref{Cbtdu}.
We find that 
\begin{equation}\label{Tonbou}
\mathcal{T} \rightarrow \infty
\end{equation}
if the curve converges to a point on the right boundary (i.e., $K = \sqrt{3 \Lambda}$)
or on the lower boundary (including the cusp).
However, if the curve converges to a point represented by $\rem \in (\rk,\infty)$ on the left boundary,
then $\mathcal{T}$ converges along this curve according to
\begin{equation}\label{Tem}
\mathcal{T} \rightarrow  \sqrt{2} \,\pi \:\Big( V(\rem) \, \frac{d^2 D}{d r^2}(\rem) \Big)^{-1/2} =: \Tem(\rem)\:,
\end{equation}
where the values of $K$ and $C$ that enter $D(r)$ are $\Kem(\rem)$ and $\Cem(\rem)$, cf.~\eqref{KCrconst}; 
see appendix~\ref{Tonboundary} for the proofs.
The function $\Tem(\rem)$ in~\eqref{Tem} is strictly monotonically decreasing in \mbox{$\rem \in (\rk,\infty)$}.
Moreover, $\Tem(\rem) \rightarrow \infty$ as $\rem \searrow \rk$ and
\begin{equation}\label{asyTu}
\mathcal{T}_m(\rem) \searrow \sqrt{\frac{3}{\Lambda}}\;\pi \qquad\quad 
(\rem \rightarrow \infty)\:.
\end{equation}
For the proof we refer to the appendix.

Based on~\eqref{Tem} and~\eqref{asyTu} we are able to complete the proof of the theorems.
Consider an integral curve $(K,C)(\tau)$
of the oriented direction field on $\mathscr{KC}_1$, see Fig.~\ref{KC1flowN}; 
we have $\mathcal{T} \equiv \mathrm{const}$ along the curve.
Since $\dot{K}>0$, the $\alpha$-limit set of the curve
must be contained in the left boundary and/or the lower boundary of $\mathscr{KC}_1$. 
However, $\mathcal{T} = \infty$ on the lower boundary (including the cusp),
hence the curve must emerge from the left boundary.
Since $\Tem(\rem)$ is strictly monotone, the curve must emerge from one point on the left boundary,
which is determined by solving the equation 
\begin{equation}\label{pointonCu}
V(r_{\mathrm{m}})\frac{d^2 D}{d r^2}(\rem) = 2 \pi^2 \mathcal{T}^2
\end{equation}
for $r_{\mathrm{m}}>\rk$, cf.~\eqref{Tem};
the associated values of $K = \Kem(\rem)$ and $C = \Cem(\rem)$ are then given through~\eqref{KCrconst}.
Since, conversely, each point on the left boundary gives rise to exactly
one integral curve of the oriented direction field on $\mathscr{KC}_1$,
we find a one-to-one correspondence between integral curves of the direction
field and points on the left boundary.

As a consequence, the range of $\mathcal{T}$ on $\mathscr{KC}_1$
coincides with the range of $\Tem$ on the left boundary.
Taking~\eqref{asyTu} into account we therefore conclude that 
$\mathcal{T}  \geq \sqrt{3}\Lambda^{-1/2}\pi$ on $\mathscr{KC}_1$
and thus on every integral curve of the direction field.
Hence, invoking Proposition~\ref{KC1embedprop} we have established Theorem~\ref{CdSThm1}.

From each point on the left boundary an integral curve emerges. The monotonicity of 
$\mathcal{T} = \Tem$ along this boundary in combination with~\eqref{asyTu}
implies that,
for a given $\mathrm{T} >  \sqrt{3}\Lambda^{-1/2}\pi$,
the equation $\mathcal{T} = \mathrm{T}$ distinguishes a unique integral curve of 
the direction field in $\mathscr{KC}_1$.
Consider a Kantowski-Sachs spacetime FC${}_\pm$[T]
with $\mathrm{T} >  \sqrt{3}\Lambda^{-1/2}\pi$.
The integral curve $\mathcal{T} = \mathrm{T}$ represents
a unique (reflection symmetric) slicing $\mathcal{S}(\tau)$ ($=\mathcal{S}_1(\tau)$)
of compact CMC hypersurfaces in FC${}_\pm$[T].
There is more, however. 
The equation $\mathcal{T} = n^{-1} \mathrm{T}$ defines a unique
integral curve in $\mathscr{KC}_1$ for each $\mathbb{N}\ni n \leq n^\star$, cf.~Theorem~\ref{CdSThm}.
Each of these integral curves corresponds
to a slicing $\mathcal{S}_n(\tau)$ in FC${}_\pm$[T], whose hypersurfaces
are isometric to the $n$-fold covering of the original data sets generated by
$(K,C)(\tau)$, cf.~Prop.~\ref{KC1embedprop}.
Accordingly, there exist $n$ slicings $\mathcal{S}_n(\tau)$, $n=1\ldots n^\star$,
in FC${}_\pm$[T].
This proves the remaining claim of Theorem~\ref{CdSThm}. \proofend

\section{Properties of the slicings}
\label{properties}

Theorem~\ref{CdSThm} establishes the existence of $n=1,\ldots,n^\star$
compact CMC slicings $\mathcal{S}_n(\tau)$, different from the $r= \rem = \mathrm{const}$ slicing,
in Kantowski-Sachs spacetimes FC${}_\pm$[T] or PC${}_\pm$[T] with
sufficiently large $\mathrm{T}$.

In the proof of Theorem~\ref{CdSThm} in section~\ref{Sec:CdST}
we have shown that $\mathcal{S}_n(\tau)$ is represented in $\mathscr{KC}_1$ by
the curve $\tau \mapsto (K,C)(\tau)$ determined by the equation $\mathcal{T} = n^{-1} \mathrm{T}$
(which is a maximally extended integral curve of the oriented direction field of Fig.~\ref{KC1flowN}).
The \textit{asymptotic properties} of the slicing $\mathcal{S}_n(\tau)$ 
are the following:

Toward the future, $K(\tau) \nearrow \sqrt{3\Lambda}$, i.e., 
the mean curvature increases monotonically and converges
to $\sqrt{3\Lambda}$; simultaneously, $C(\tau) \rightarrow\infty$.
These results follow straightforwardly from the discussion of section~\ref{Sec:CdST}.
In particular, we find that $r\rightarrow \infty$ along $\mathcal{S}_n(\tau)$
as $\tau\rightarrow \infty$ (or, if we use a different time gauge, 
as $\tau$ approaches the finite supremum of its range),
which follows from the fact that $r_{\min} \rightarrow \infty$ as $C\rightarrow \infty$
in $\mathscr{KC}_1$.

Toward the past, the slicing $\mathcal{S}_n(\tau)$ converges to a limit hypersurface.
Let $\rem$ be the solution of $\Tem(\rem) = n^{-1} \mathrm{T}$, i.e., $\rem$ is the solution
of
\begin{equation}
V(r_{\mathrm{m}})\frac{d^2 D}{d r^2}(\rem) = 2 \pi^2 n^{-2}\, \mathrm{T}^2 \:,
\end{equation}
cf.~\eqref{Tem} and~\eqref{pointonCu}.
Then $K(\tau) \searrow \Kem(\rem)$ and $C(\tau)\searrow \Cem(\rem)$
as $\tau \searrow \tauem$ (where $\tauem$ denotes the infimum
of the maximal range of $\tau$).
These statements follow straightforwardly from the discussion of section~\ref{Sec:CdST}.
In particular, since $r_{\min}(\tau) \rightarrow \rem$ and $r_{\max}(\tau)\rightarrow \rem$,
see appendix~\ref{Tonboundary}, we find that the slicing $\mathcal{S}_n(\tau)$ converges
to the hypersurface of constant $r = \rem$ as $\tau \searrow \tauem$.

A natural question in connection with the slicing $\mathcal{S}_n(\tau)$ 
concerns its limit hypersurface \mbox{$r = \rem$}.
Let us first note that for $\mathcal{S}_n(\tau)$ we have  
$\dot{K}(\tau)\rightarrow 0$ as $\tau\rightarrow \tauem$, where we assume 
a gauge of $\tau$ such that $\alpha$ remains bounded in the limit.
To see this, consider the defining equation~\eqref{Cdotonemin} for the direction field.
Since $(d D/d r)(r_{\min})$ converges to zero as $\tau\rightarrow\tauem$,
because \mbox{$r_{\min} = r_{\min}(\tau) \rightarrow \rem$},
$(\dot{K} r_{\min}/3 -\dot{C}/r_{\min}^2) \rightarrow 0$ in this limit.
An additional argument then shows that $\dot{K}\rightarrow 0$ and $\dot{C}\rightarrow 0$
separately as $\tau\rightarrow \tauem$.
(Suppose $\dot{K}\not\rightarrow 0$; then $\dot{C}\not\rightarrow 0$ and
$\dot{C} \sim (r_{\min}^3/3) \dot{K}$ as $\tau\rightarrow \tauem$.
This implies that $(\dot{K},\dot{C})$ is tangential to
the left boundary of $\mathscr{KC}_1$, since $d \Cem/d\rem = (\rem^3/3) (d\Kem/d\rem)$
by~\eqref{KCrconst}. This is false, however; the oriented direction
field is transversal to the left boundary, which can be straightforwardly
deduced from the behavior of the function $\mathcal{T}$ on the boundary
and in a small neighborhood thereof. We simply recall that $\mathcal{T}$ ($ = \Tem(\rem)$)
is strictly monotone along the boundary, while $\mathcal{T} = \mathrm{const}$
along the integral curves of the direction field.)

Let us now consider the lapse equation and its solutions 
on the hypersurface $r=r_{\mathrm{m}}$, i.e., for $(K,C) =\big(\Kem(\rem),\Cem(\rem)\big)$.
First, recall that the metric on $r=r_{\mathrm{m}}$
is given by $d l^2 + r_{\mathrm{m}}^2 \metricS$; the variable 
$l$ is related to $t$ by $l = t \sqrt{-V(r_{\mathrm{m}})}$. (This follows directly from~\eqref{KSSdSmetric}
or from~\eqref{tofl} with~\eqref{Vremprop}.)
Since $r=r_{\mathrm{m}}$ is regarded as a hypersurface in FC${}_\pm$[T] or PC${}_\pm$[T], 
$l=-L$ is identified with $l=L$, where 
$L =\sqrt{-V(r_{\mathrm{m}})}\, \mathrm{T}$.
The lapse equation $\Delta_g \alpha + a \alpha = \dot{K}$ specializes to 
\begin{equation}\label{lapseCu}
\alpha^{\prime\prime} + a \,\alpha = \dot{K} \:,
\end{equation}
where $a = \Lambda - K^2/3 - 6 C^2/\rem^6 = \mathrm{const}$. 
Standard algebraic manipulations 
show that $a>0$ for $(K,C) =\big(\Kem(\rem),\Cem(\rem)\big)$, for all $\rem \in (\rk,\infty)$,
and $a\rightarrow 0$ as $K\rightarrow \Kk$ (i.e., $\rem \rightarrow \rk$).
In the case $\dot{K}=0$ the general even solution of~\eqref{lapseCu} is
\begin{equation}\label{lapseCu0}
\alpha(l) = -A \cos \left(\sqrt{a} \, l\right) 
\end{equation}
with $A \in\mathbb{R}$. (In the case $\dot{K} \neq 0$ we obtain $\alpha(l) = \dot{K}/a - A \cos(\sqrt{a}\, l)$.)
Intuitively speaking, the lapse function~\eqref{lapseCu0} connects $r=r_{\mathrm{m}}$ with 
a infinitesimally neighboring CMC hypersurface in FC${}_\pm$[T] or PC${}_\pm$[T], if and only if
$\sqrt{a}\, L =  \pi n$ for some $n\in\mathbb{N}$, i.e., if
\mbox{$n \pi a^{-1/2} (-V(r_{\mathrm{m}}))^{-1/2} = \mathrm{T}$}.
Using that $a = -(1/2) (d^2 D/d r^2)(r_{\mathrm{m}})$, it is straightforward to see that
this condition is equivalent to $\mathrm{T} = n \Tem(\rem)$, cf.~\eqref{Tem}.
This is in prefect accord with the previous results and we conclude that
the slicing $\mathcal{S}_n(\tau)$, $\tau \in (\tauem,\infty)$, can be smoothly
extended to a slicing $\tau \in [\tauem,\infty)$, which also includes $r=r_{\mathrm{m}}$.
In addition, the slicing $\mathcal{S}_n(\tau)$, $\tau \in [\tauem,\infty)$, can be joined continuously 
to the slicing of $r=\mathrm{const}$ hypersurfaces at $r=r_{\mathrm{m}}$.
However, the junction is not smooth, because 
for the slicing of $r=\mathrm{const}$ hypersurfaces we have $\dot{K}\neq 0$ and 
the lapse function is $\alpha=\dot{K}/a = \mathrm{const}(\tau)\neq 0$,
while $\dot{K} = 0$ and~\eqref{lapseCu0} holds
for $\mathcal{S}_n(\tau)$ at $\tau = \tauem$.

It remains to  investigate whether the slicings $\mathcal{S}_n(\tau)$ are in fact foliations.
In analogy to \mbox{\cite[App.~C]{Beig/Heinzle:2005}} we may define 
$\mathscr{KC}_{1+}$ as the set of all $(K,C)\in\mathscr{KC}_1$ such that 
the lapse function $\alpha$ is positive. (Then the slicing
$\mathcal{S}_n(\tau)$ is a foliation for those values of $\tau$ for which 
the integral curve $(K,C)(\tau)$ associated with $\mathcal{S}_n(\tau)$ 
is in $\mathscr{KC}_{1+}$.)
Recall the parameter space $\mathscr{KC}_1$ from Figs.~\ref{Cbtdu} and~\ref{KC01}.
We can prove that there exists a neighborhood $\mathcal{U}_{\mathrm{left}}$ in $\overline{\mathscr{KC}_1}$ of 
the left boundary and a neighborhood $\mathcal{U}_{\mathrm{lower}}$ of 
the lower boundary of $\mathscr{KC}_1$
such that $\mathcal{U}_{\mathrm{left}} \cap \mathscr{KC}_{1+}=\emptyset$ 
and $\mathcal{U}_{\mathrm{lower}} \cap \mathscr{KC}_{1+}=\emptyset$. 
The former statement follows immediately from~\eqref{lapseCu0},
the proof of the latter statement is analogous to the proof of~\cite[Prop.~C.2]{Beig/Heinzle:2005}.
Furthermore, there exists a neighborhood $\mathcal{U}_{\mathrm{right}}$ in $\overline{\mathscr{KC}_1}$
of the straight line $K=\sqrt{3 \Lambda}$, which is the right boundary of $\mathscr{KC}_1$,
such that $\mathcal{U}_{\mathrm{right}} \cap \mathscr{KC}_1 \subseteq \mathscr{KC}_{1+}$.
(To see this we exploit the fact that $\alpha > 0$ as $K\rightarrow \sqrt{3\Lambda}$, 
which is intimately connected with the fact that
the CMC hypersurfaces with $K = \sqrt{3 \Lambda}$, which are not compact,
form a foliation; for the details we refer to upcoming work. ).
Numerical investigations suggest that $\mathscr{KC}_{1+}$ is a connected set, whose
minimal value of $K$ is strictly greater than $\Kk$. For large $C$,
the boundary $\partial(\mathscr{KC}_{1+})$ approximates the
left boundary of $\mathscr{KC}_1$;
this leads to the fact that every integral curve lies in
$\mathscr{KC}_{1+}$ for sufficiently large $\tau$. Expressed in terms
of the associated slicing in FC${}_\pm$[T]:
$\mathcal{S}_n(\tau)$ is indeed a foliation for sufficiently large $\tau$.

A concluding remark:
Compact CMC hypersurfaces and slicings in spacetimes that satisfy the timelike convergence condition
are a (comparatively) straightforward matter.
Spacetimes that do not satisfy the timelike convergence condition, on the other hand,
may exhibit a rich structure of compact CMC hypersurfaces
and CMC slicings.
The vacuum Kantowski-Sachs spacetimes with positive cosmological constant
are a fine example to illustrate this fact and to provide
indications about what to expect
of CMC slicings in more general spacetimes.

\begin{appendix}

\section{The function $\pmb{\mathcal{T}}$ on $\pmb{\mathscr{KC}_1}$}
\label{Tonboundary}

This appendix is concerned with an analysis of the function $\mathcal{T}$ on
boundary of $\mathscr{KC}_1$ and the proof of the statements of section~\ref{Sec:CdST}
in connection with~\eqref{Tonbou},~\eqref{Tem}, and~\eqref{asyTu},
which are essential for the proof of Theorem~\ref{CdSThm1} and Theorem~\ref{CdSThm}.

The boundary of the parameter space $\mathscr{KC}_1$ consists of 
(a piece of) the straight line $K = \sqrt{3 \Lambda}$,
the curve $(K,C)(\rem)$, $\rem \in (r_\ast, \rk]$, which corresponds
to the lower boudary and includes the cusp,
and the curve $(K,C)(\rem)$, $\rem \in (\rk, \infty)$,
which is the left boundary, see~\eqref{KCrconst} and Figs.~\ref{Cbtdu} and~\ref{KC01}.
In this context we have 
\[
r_\ast = \frac{1}{\sqrt{\Lambda}}\:\Big( 1 + \sqrt{1 - 3 \sqrt{\Lambda} M} \Big)\:,
\]
which follows by setting $K=\sqrt{3\Lambda}$ in~\eqref{rprime2}
and using that $D(r_\ast) = (d D/d r)(r_\ast) = 0$. 

Let us begin by proving~\eqref{Tonbou}.
Consider a curve $\sigma \mapsto (K, C)(\sigma) = (K_\sigma, C_\sigma)$ in $\mathscr{KC}_1$ that converges, 
as $\sigma \searrow 0$, to a point $(K_0,C_0)$ on the lower boundary that
is represented by $\rem \in (r_\ast, \rk)$.
Let $D_\sigma(r)$ be the function defined in~\eqref{rprime2} with $K = K_\sigma$ and $C= C_\sigma$;
its profile is that of Fig.~\ref{Deltaborderline}(a). As $\sigma \rightarrow 0$, however,
$D_\sigma(r) \rightarrow D_0(r)$, where $D_0(r)$ is of the form depicted in Fig,~\ref{Deltaborderline}(b)
(since $(d^2 D_0/d r^2)(\rem) > 0$ because of the assumption $\rem < \rk$).

From~\eqref{Tdef} we have
\begin{equation}\label{Tsig}
\mathcal{T}_\sigma = \mathcal{T}(K_\sigma, C_\sigma)  
= \Big|\int\limits_{r_{\min, \sigma}}^{r_{\max,\sigma}} V^{-1}(r) D_\sigma^{-1/2}(r)
\:\Big(\frac{K_\sigma r}{3} -\frac{C_\sigma}{r^2}\Big)  d r\Big|\:,
\end{equation}
where $r_{\min,\sigma}$ and $r_{\max,\sigma}$ are the relevant zeros 
of $D_{\sigma}(r)$, see Fig.~\ref{Deltaborderline}(a).
As $\sigma\rightarrow 0$, we have $r_{\min,\sigma}\rightarrow \rem$
and $r_{\max,\sigma} \rightarrow r_{\max,0}$, where $r_{\max,0}$ is a simple
zero of $D_0(r)$.

Let $0 < \epsilon \ll 1$;  
we split~\eqref{Tsig} into an integral over the interval $(r_{\min,\sigma}, r_{\min,\sigma} +\epsilon)$
and another over the interval $(r_{\min,\sigma} +\epsilon, r_{\max,\sigma})$.
The latter converges to some constant as $\sigma\rightarrow 0$;
the former, however, and thus $\mathcal{T}_\sigma$ itself,
will be shown to diverge as $\sigma\rightarrow 0$.
To that end, we make the expansion 
\[
D_\sigma(r) = D'_\sigma(r_{\min,\sigma}) \,(r - r_{\min,\sigma}) +  \textfrac{1}{2}\, D^{\prime\prime}_\sigma(r_{\min,\sigma}) \:(r-r_{\min,\sigma})^2 +
O\big((r-r_{\min,\sigma})^3\big) \:,
\]
as $r\rightarrow r_{\min,\sigma}$, where we note that $D'_\sigma(r_{\min,\sigma}) \rightarrow 0$ and $D^{\prime\prime}_\sigma(r_{\min,\sigma}) \rightarrow \mathrm{const} > 0$
as $\sigma\rightarrow 0$. In this context, a prime denotes the derivative w.r.t.\ the argument, i.e., $r$.
We then find
\[
D_\sigma^{-1/2}(r) = \Big( D'_\sigma(r_{\min,\sigma}) \,(r - r_{\min,\sigma}) +  \textfrac{1}{2}\, D^{\prime\prime}_\sigma(r_{\min,\sigma}) \:(r-r_{\min,\sigma})^2 \Big)^{-1/2}
\,+ O(1) 
\]
as $r\rightarrow r_{\min,\sigma}$, where $O(1)$ is independent of $\sigma$.
In addition,
\[
V^{-1}(r) \:\Big(\frac{K_\sigma r}{3} -\frac{C_\sigma}{r^2}\Big) = 
V^{-1}(r_{\min,\sigma}) \:\Big(\frac{K_\sigma r_{\min,\sigma}}{3} -\frac{C_\sigma}{r_{\min,\sigma}^2}\Big) + O\big( (r -r_{\min,\sigma}) \big) \:.
\]
Then integration yields
\begin{align*}
& \Big|\int\limits_{r_{\min, \sigma}}^{r_{\min,\sigma}+\epsilon} V^{-1} D_\sigma^{-1/2}
\:\Big(\frac{K_\sigma r}{3} -\frac{C_\sigma}{r^2}\Big)  d r\Big| \: = \\
 & \qquad = |V(r_{\min,\sigma})|^{-1} \:\Big(\frac{K_\sigma r_{\min,\sigma}}{3} -\frac{C_\sigma}{r_{\min,\sigma}^2}\Big)\:
\frac{2\sqrt{2}}{\sqrt{D^{\prime\prime}_\sigma(r_{\min,\sigma})}} \:\arsinh\sqrt{\frac{\epsilon}{2}\:\frac{D^{\prime\prime}_\sigma(r_{\min,\sigma})}{ D'_\sigma(r_{\min,\sigma})}}
+ O(\epsilon) \:,
\end{align*}
where $O(\epsilon)$ is independent of $\sigma$.
Letting $\sigma\rightarrow 0$ while keeping $\epsilon$ fixed, we obtain divergence,
because $D'_\sigma(r_{\min,\sigma}) \rightarrow 0$.
Accordingly, $\mathcal{T}_\sigma \rightarrow \infty$ as $\sigma\rightarrow 0$.

Likewise, we find divergence of $\mathcal{T}$ along every curve in $\mathscr{KC}_1$ that
converges to a point on the right boundary $K = \sqrt{3 \Lambda}$.
To see that, consider a curve $\sigma \mapsto (K, C)(\sigma) = (K_\sigma, C_\sigma)$ in $\mathscr{KC}_1$ that converges, 
as $\sigma \searrow 0$, to a point $(K_0,C_0) =  (\sqrt{3 \Lambda},C_0)$ on $\partial\mathscr{KC}_1$.
The profile of $D_\sigma(r)$ is that of Fig.~\ref{Deltaborderline}(a). As $\sigma \rightarrow 0$, 
we have $D_\sigma(r) \rightarrow D_0(r)$, where
\[
D_0(r) = 1 - \frac{2 M}{r} -\frac{2 \sqrt{3\Lambda}\, C_0}{3 r} + \frac{C_0^2}{r^4} \:,
\]
which entails that the zero $r_{\min,\sigma}$ of $D_\sigma(r)$ converges to $r_{\min,0}$, which is a zero
of $D_0(r)$; however, $r_{\max,\sigma} \rightarrow \infty$ as $\sigma \rightarrow 0$.

For sufficiently small $\sigma$, there exists a uniform bound for $D_\sigma(r)$  
on $[r_{\min,\sigma}, r_{\max,\sigma}]$; furthermore, 
$K_\sigma/3 - C_\sigma/r^3$ is positive on $\overline{\mathscr{KC}_1}$ and thus 
uniformly bounded from below.
Therefore,
\begin{align*}
\mathcal{T}_\sigma & = 
\Big|\int\limits_{r_{\min, \sigma}}^{r_{\max,\sigma}} V^{-1}(r) D_\sigma^{-1/2}(r)
\:\Big(\frac{K_\sigma r}{3} -\frac{C_\sigma}{r^2}\Big)  d r\Big| \\
& \geq 
\mathrm{const}\:\int\limits_{r_{\min, \sigma}}^{r_{\max,\sigma}} |V^{-1}(r)|\: r\, d r
\geq
\mathrm{const}\:\int\limits_{r_{\min, \sigma}}^{r_{\max,\sigma}} r^{-1} d r =
\mathrm{const} \; \log r \:\Big|_{r_{\min, \sigma}}^{r_{\max,\sigma}} \:,
\end{align*}
and, since $r_{\max,\sigma} \rightarrow \infty$ as $\sigma \rightarrow 0$,
we obtain $\mathcal{T}_\sigma\rightarrow \infty$ in this limit.
This concludes the proof of~\eqref{Tonbou}.

Let us proceed by proving~\eqref{Tem}.
Consider a curve $\sigma \mapsto (K, C)(\sigma) = (K_\sigma, C_\sigma)$ in $\mathscr{KC}_1$ that converges, 
as $\sigma \searrow 0$, to a point $(K_0,C_0)$ on the left boundary that
is represented by \mbox{$\rem \in (\rk,\infty)$}.
As $\sigma \rightarrow 0$, 
$D_\sigma(r) \rightarrow D_0(r)$, where $D_0(r)$ is of the form depicted in Fig.~\ref{Deltaborderline}(c).
Let $r_{+,\sigma}$ be defined as the maximum of $D_\sigma(r)$, see Fig.~\ref{Deltaborderline}(a);
obviously, the three values $r_{\min,\sigma} < r_{+,\sigma} < r_{\max,\sigma}$ converge to $\rem$
as $\sigma\rightarrow 0$.

Set $\tilde{r}_{+,\sigma} = r_{+,\sigma}$ (for purely aesthetic reasons) and consider the function
\begin{equation}
\twoDelta_\sigma(\tilde{r}) = D_\sigma(\tilde{r}_{+,\sigma})  + D^{\prime\prime}_\sigma(\tilde{r}_{+,\sigma}) \,\frac{(\tilde{r}-\tilde{r}_{+,\sigma})^2}{2}\:,
\end{equation}
whose zeros we denote by $\tilde{r}_{\min,\sigma}$ and $\tilde{r}_{\max,\sigma}$.
Obviously, there exists a diffeomorphism $r(\tilde{r})$ such that
$D_\sigma\big(r(\tilde{r})\big) = \tilde{D}_\sigma(\tilde{r})$ in the interval $[\tilde{r}_{\min,\sigma}, \tilde{r}_{\max,\sigma}]$;
we have $r(\tilde{r}_{\min,\sigma}) = r_{\min,\sigma}$, \mbox{$r(\tilde{r}_{+,\sigma}) = r_{+,\sigma}$}, and $r(\tilde{r}_{\max,\sigma}) = r_{\max,\sigma}$.
We find
\begin{align*}
\int\limits_{r_{\min,\sigma}}^{r_{\max,\sigma}} \frac{1}{\sqrt{D_\sigma(r)}} \:d r & =
\int\limits_{\tilde{r}_{\min,\sigma}}^{\tilde{r}_{\max,\sigma}} \frac{1}{\sqrt{\twoDelta_\sigma(\tilde{r})}} \,\frac{d r(\tilde{r})}{d\tilde{r}} \:d \tilde{r} =
\int\limits_{\tilde{r}_{\min,\sigma}}^{\tilde{r}_{\max,\sigma}} \frac{1}{\sqrt{\twoDelta_\sigma(\tilde{r})}} \, 
\frac{\twoDelta^\prime_\sigma(\tilde{r})}{D^\prime_\sigma\big(r(\tilde{r})\big)} \:d \tilde{r} \:,
\intertext{where the prime denotes differentiation w.r.t.\ the argument.
Since the quotient of the derivatives is of the form
$1 + o(1)$ as $\sigma\rightarrow 0$, by de l'Hospital's rule, we further obtain}
\int\limits_{r_{\min,\sigma}}^{r_{\max,\sigma}} \frac{1}{\sqrt{D_\sigma(r)}} \:d r & =
\int\limits_{\tilde{r}_{\min,\sigma}}^{\tilde{r}_{\max,\sigma}} \frac{1}{\sqrt{\twoDelta_\sigma(\tilde{r})}}\:d \tilde{r}\, +\, o(1)\:.
\end{align*}
It is straightforward to calculate the integral on the r.h.\ side: We obtain 
$\sqrt{2} \,\pi\,\sqrt{-D^{\prime\prime}_\sigma(r_+)^{-1}}$.
From this result, the claim~\eqref{Tem}, i.e.,
\[
\mathcal{T}_\sigma \,\rightarrow\, \Tem(\rem) := \sqrt{2} \,\pi \:\Big( V(\rem) \, \frac{d^2 D_0}{d r^2}(\rem) \Big)^{-1/2} \quad (\sigma\rightarrow 0)\:,
\]
follows immediately by noting that 
\begin{equation}\label{Vremprop}
|V^{-1}(\rem)| \Big( \frac{K_0 \rem}{3} - \frac{C_0}{\rem^2} \Big)  = 
|V^{-1}(\rem)| \Big( \frac{\Kem(\rem) \rem}{3} - \frac{\Cem(\rem)}{\rem^2} \Big)  = 
\sqrt{|V^{-1}(\rem)|} \:,
\end{equation}
which is because $D_0(\rem) = 0$. 

It remains to prove the claims in connection with~\eqref{asyTu}.
Let $D(r)$ be the function of~\eqref{rprime2} associated with $(K,C) = (\Kem(\rem),\Cem(\rem))$.
First, to establish strict monotonicity of $\Tem(\rem)$, we consider $V(r_{\mathrm{m}}) D^{\prime\prime}(r_{\mathrm{m}})$ and
compute
\[
\frac{\partial}{\partial \rem}\: \left[V(\rem)D^{\prime\prime}(\rem)\:\right] =
8 \rem^{-3}\, \left(1- \frac{3 M}{\rem}\right)^2\: > 0 \:;
\]
strict inequality holds because 
$\rem > \rk > r_{\mathrm{c}} > 3 M$ ($ > r_{\mathrm{b}}$).
(Note that the computation involves derivatives of~\eqref{KCrconst} w.r.t.\ $\rem$,
since $K = \Kem(\rem)$ and $C = \Cem(\rem)$ in $D(r)$.)
Second, $\Tem(\rem) \rightarrow \infty$ as $\rem \searrow \rk$, since
$D^{\prime\prime}(\rem) \rightarrow 0$ as $\rem$ approaches
the `cusp' $\rk$.
Third, to show~\eqref{asyTu} we compute
\[
D^{\prime\prime}(\rem) = \frac{2}{3}\: V(\rem)^{-1} \, \Big( \Lambda - 3 \rem^{-4} \big[ (\rem-3 M)^2 + \rem (\rem- 2M) \big] \Big) \:,
\]
hence
\[
\Tem(\rem) =  \sqrt{2} \,\pi \:\Big( V(\rem) \, D^{\prime\prime}(\rem) \Big)^{-1/2}
= \sqrt{2} \,\pi\: \sqrt{\frac{3}{2 \Lambda}} + O(\rem^{-2}) = \sqrt{\frac{3}{\Lambda}}\: \pi + O(\rem^{-2}) 
\]
as $\rem \rightarrow \infty$, and~\eqref{asyTu} is proved.

\end{appendix}


\bibliographystyle{plain}


\end{document}